\documentclass[aip,jcp,twocolumn,10pt]{revtex4-1}
\usepackage{dcolumn}
\usepackage{bm}
\usepackage{color}
\usepackage[english]{babel}
\usepackage{float}
\usepackage{amsmath}
\usepackage{graphicx}
\usepackage{times}   %% Times Roman font
\usepackage{esint}
\usepackage{subfigure}
\usepackage{enumerate}
\usepackage{latexsym}
\usepackage[unicode=true,pdfusetitle,
 bookmarks=false,colorlinks=true,citecolor=blue,urlcolor=blue,linkcolor=red]{hyperref}

\makeatletter

\@ifundefined{textcolor}{}
{%
 \definecolor{BLACK}{gray}{0}
 \definecolor{WHITE}{gray}{1}
 \definecolor{RED}{rgb}{1,0,0}
 \definecolor{GREEN}{rgb}{0,1,0}
 \definecolor{BLUE}{rgb}{0,0,1}
 \definecolor{CYAN}{cmyk}{1,0,0,0}
 \definecolor{MAGENTA}{cmyk}{0,1,0,0}
 \definecolor{YELLOW}{cmyk}{0,0,1,0}
}

\@ifundefined{date}{}{\date{}}
\AtBeginDocument{
  
}
\makeatother

\newcommand{\SAVE}[1]{}
\newcommand{\newt}[1]{{#1}}

\begin{document}
\renewcommand\abstractname{}

\title{Density-matrix based determination of low-energy model Hamiltonians from {\it ab initio} wavefunctions}
\author{Hitesh J. Changlani}
\author{Huihuo Zheng}
\author{Lucas K. Wagner}
\affiliation{Department of Physics, University of Illinois at Urbana-Champaign, 1110 West Green St, Urbana IL 61801, USA}
\date{\today}

\begin{abstract}
We propose a way of obtaining effective low energy Hubbard-like model 
Hamiltonians from {\it ab initio} quantum Monte Carlo (QMC) 
calculations for molecular and extended systems. The Hamiltonian parameters 
are fit to best match the {\it ab initio} two-body density matrices 
and energies of the ground and excited states, and thus we refer to the method 
as {\it ab initio} density matrix based downfolding (AIDMD).  
For benzene (a finite system), we find 
good agreement with experimentally available energy gaps 
without using any experimental inputs. 
For graphene, a two dimensional solid (extended system) 
with periodic boundary conditions, we find the effective on-site 
Hubbard $U^{*}/t$ to be $1.3 \pm 0.2$, comparable to 
a recent estimate based on the 
\newt{constrained random phase approximation}. 
For molecules, such parameterizations enable calculation of 
excited states that are usually not accessible within ground state approaches. 
For solids, the effective Hamiltonian enables large-scale calculations 
using techniques designed for lattice models.  
\end{abstract}

\maketitle
\section{Introduction}
\label{sec:introduction}

The reliable simulation of systems for which the large-scale physics is not well-approximated
by a non-interacting model, is a major challenge in physics, chemistry, and materials science.
These systems appear to require a multi-scale approach 
in which the effective interactions between electrons at a small distance scale are determined, 
which then leads to a coarse-grained description of emergent correlated physics.
This reduction of the Hilbert space is often known as "downfolding".
In strongly-correlated systems, the correct effective Hamiltonian 
is strongly dependent on material-specific properties, 
motivating the need for a generic accurate method to determine it. 

One can loosely categorize downfolding techniques into two strategies.
The first strategy is based on performing {\it ab initio} calculations and 
then matching them state by state to the effective model. Alternately, 
some approaches employ a model for the 
screening of Coulomb interactions, for which the {\it ab initio} single particle wavefunctions 
provide the relevant inputs. For the purposes of this manuscript, 
the umbrella term we use for these strategies is "fitting". 
Techniques that fall into this class include the constrained 
density functional theory~\cite{Pavirini,Dasgupta}, 
the constrained random phase approximation (cRPA)~\cite{Aryasetiawan}, 
fitting spin models to energies~\cite{Valenti_kagome,Spaldin}, 
and efforts by one of us~\cite{Wagner_JCP} using reduced density matrices of quantum 
Monte Carlo (QMC) calculations. The second class is based on 
L\"owdin downfolding~\cite{Freed,Zhou_Ceperley,Tenno} 
and canonical transformation theory~\cite{Glazek_Wilson,Wegner,White_CT,Yanai_CT,Watson_Chan}, 
which involves a sequence of unitary transformations on the 
bare Hamiltonian, chosen in a way that minimize the size of the matrix elements 
connecting the relevant low energy (valence) space to the high energy one.

Downfolding by fitting has the advantage that it 
is conceptually straightforward to perform, although it demands an 
{\it a priori} parameterization of the effective Hamiltonian. 
The methods have been applied to complex bulk systems~\cite{Imada1,Imada2,Arya1,Arya2,Scriven,Wehling_graphene}, but it 
is only recently that their 
accuracy is being rigorously checked~\cite{RPA_Troyer}. 
On the other hand, canonical transformations do not need 
such parameterizations and can discover the relevant terms in an automated way. 
However, their application to complex materials remains to be carried out 
and tested.

Here we propose a scheme which aims to capture 
the "best of both worlds". On the one hand, we 
retain the simplicity of fitting and on the other we 
use information from accurate many-body wavefunctions to 
determine which terms are important. The deviations 
between \newt{the} {\it ab initio} and model properties 
allows us to assess the quality of the resultant model and to discover 
relevant physics from the calculation. Simultaneously, the method cannot 
depend too much on the quality of the {\it ab initio} solution 
because of the inherent limitations of accuracy, especially for 
very big system sizes.

\begin{figure}[htpb]
\centering
\includegraphics[width=1\linewidth]{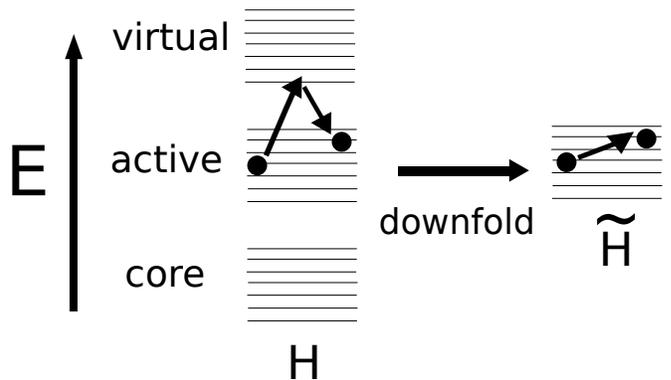}
\caption{\newt{Schematic for downfolding. The full Hamiltonian H is defined in the space of active (partially occupied), 
core (mostly occupied) and virtual (mostly unoccupied) orbitals. These orbitals 
have been arranged according to their energy (E) in the figure. The objective is to map the physics 
of the original system to that of the effective one, defined only in the active space, 
with Hamiltonian $\tilde{H}$.}}
\label{fig:effham} 
\end{figure}

Once an effective model Hamiltonian in the reduced Hilbert space 
is obtained, as is depicted in Fig.~\ref{fig:effham}, 
it can be used to perform a calculation on a 
larger system using techniques designed for 
lattice models~\cite{dmrg_white,tps_nishino,Vidal_MERA,TPS_review,Changlani_CPS,
Neuscamman_CPS,mezzacapo,Marti,DMET_2012,Chen_DMET,Sandvik_loops,Blankenbecler,Alavi_FCIQMC,SQMC}. 
This multi-step modeling procedure is needed since the {\it ab initio} 
calculations for a given system size are, in general, 
computationally more expensive than the equivalent lattice calculations. 
Large sizes are crucial to study finite size effects, and in turn 
theoretically establish the presence of a phase. 
In addition, excited states and dynamical correlation functions have traditionally 
been difficult in {\it ab initio} approaches, 
but have seen progress for lattice model methods~\cite{dynamical_dmrg, Daley_tDMRG, White_tDMRG}.

In this paper, we demonstrate a downfolding method that uses data from {\it ab initio} QMC 
techniques to derive an effective coarse-grained Hamiltonian.
This method, which we call non-eigenstate {\it ab initio} density matrix downfolding (N-AIDMD), 
uses many-body simulations of non-eigenstates to fit an effective low-energy Hamiltonian.
We demonstrate that the method can use wavefunctions of medium quality to derive highly accurate effective Hamiltonians. 
\newt{After demonstrating a simple example,} we downfold benzene from a 30-electron problem to a 6-electron 
one and show that the resulting Hamiltonian reproduces the experimental spectrum well. 
We also show that the method works for extended systems, by applying it to graphene.

\section{Methods}
In the present section we discuss the methods we used to generate our 
{\it ab initio} data. 
While most of our discussion is specific to QMC, the quantities used can also be calculated in 
almost any other wavefunction-based quantum chemistry method. 

\subsection{Variational and Fixed Node Diffusion Quantum Monte Carlo}
{\it Ab initio} QMC comprises of a suite of methods that efficiently sample 
the phase space of $N$ electrons each moving in 3-dimensional real space. 
When the wavefunction as a function of $3N$ coordinates is known, the phase space
can be sampled with variational Monte Carlo (VMC) using Metropolis algorithms.
For ground state calculations, the Diffusion Monte Carlo (DMC) method, 
based on imaginary time evolution of the Schroedinger equation, is formally exact but in practice  
limited by the numerical sign problem. This problem ceases to exist if one knew the exact location 
of the nodes (zeroes) of the many-body wavefunction. Thus, the optimal 
strategy for very accurate calculations is 
to generate a good trial wavefunction, and optimize its parameters to 
minimize its variational energy. Then, we use this wavefunction to "fix the nodes" (which may 
only approximately correspond to the exact nodes) and perform a DMC calculation under this constraint.  
This last variant is called the fixed-node DMC (FN-DMC) method and is known to be 
very accurate for a large variety of systems. While some more details are discussed here, 
we refer the interested reader to Ref.~\onlinecite{Foulkes_review} for an exhaustive 
review of concepts and applications. All the {\it ab initio} QMC calculations were carried 
out with the QWalk package~\cite{QWALK}.

A typical QMC calculation was carried out as follows. 
First, we perform DFT calculations with the B3LYP~\cite{B3LYP} or 
PBE functionals~\cite{PBE} using GAMESS~\cite{GAMESS} 
for molecules or CRYSTAL~\cite{CRYSTAL} for solids.  
The lowest energy DFT orbitals provide the Slater determinant part of the trial 
wavefunction. For molecules, a multi-determinantal wavefunction is generated by performing 
a configuration interaction with singles and doubles excitations (CISD) \newt{calculation} 
from the reference Slater determinant. 
Once this is done, a Jastrow factor \newt{$\mathcal{J}$} 
is introduced, resulting in the ansatz \newt{for the trial wavefunction $\psi_T$},
\begin{equation}
	\psi_T (r_1,r_2,....r_N) = \mathcal{J} \sum_{i} d_i D_i
\end{equation} 
\newt{where $(r_1,r_2,....r_N)$ refers to the coordinates of the electrons (the spin 
indices, being fixed, have been suppressed), 
$D_i$ are determinants and $d_i$ their corresponding coefficients.} 
When we desire eigenstates, the parameters in the Jastrow $\mathcal{J}$ and 
the coefficients $d_i$ are optimized to 
get the best possible variational energy within the ansatz chosen 
using a technique introduced by Umrigar and coworkers~\cite{Umrigar_optimization} 
with an efficient \newt{algorithm} by Clark et al.~\cite{Clark_multidet}.

The variational energy is calculated via Metropolis sampling of $|\psi_T|^{2}$, 
\begin{equation} 
 E_{VMC} \equiv \frac{ \langle \psi_T | H | \psi_T \rangle} {\langle \psi_T | \psi_T \rangle} = \frac{ \int |\psi_T ({\bf{R}})|^2 \frac{H \psi_T({\bf{R}})}{\psi_T({\bf{R}})} \;\;\; d{\bf{R}} } { \int |\psi_T ({\bf{R}})|^2 d {\bf{R}}}
\end{equation}
where $\bf{R}$ is a compact notation for the coordinates of the electrons 
and $H \psi_T({\bf{R}})/\psi_T({\bf{R}})$ is the "local energy". 
With this trial wavefunction we perform FN-DMC 
to calculate the energy using the mixed (or projected) estimator, 
\begin{equation}
E_{DMC} \equiv \frac{ \langle \psi | H | \psi_T \rangle} {\langle \psi | \psi_T \rangle} = \frac{ \int \psi_T ({\bf{R}}) \psi({\bf{R}}) \frac{H \psi_T({\bf{R}})}{\psi_T({\bf{R}})} d {\bf{R}} }{ \int \psi_T ({\bf{R}}) \psi ({\bf{R}}) d {\bf{R}}}
\end{equation}
where $\psi \equiv \exp (-\beta H) \psi_T$ is obtained by a stochastic projection 
of $\psi_T$ under the constraint that $\psi$ and $\psi_T$ have the same sign everywhere. 

We now discuss measurements in the QMC methods. For a generic operator $\hat{O}$, 
the pure (VMC) and mixed estimators are computed as, 
\begin{equation}
         \langle \hat{O} \rangle_{VMC} \equiv \frac{ \langle \psi_T | \hat{O} | \psi_T \rangle}   {\langle \psi_T | \psi_T \rangle} 
\;\;\;\; \langle \hat{O} \rangle_{mix} \equiv \frac{ \langle \psi   | \hat{O} | \psi_T   \rangle} {\langle \psi   | \psi_T \rangle}
\end{equation}
The mixed estimator of an operator is equal to the pure estimator 
in two cases; (1) when $\psi_T$ is the exact wavefunction or (2) when the 
operator $\hat{O}$ commutes with the Hamiltonian. In more general situations, 
higher accuracy can be obtained with the extrapolated estimator\cite{Ceperley_Kalos_Binder} for 
approximate eigenstates,
\begin{equation}
	\langle \hat{O} \rangle_{\text{extrap}}  = 2 \langle \hat{O} \rangle_{\text{mix}} - \langle \hat{O} \rangle_{\text{VMC}}
\end{equation}
\newt{For accurate wavefunctions}, all these estimators must approach the same value. 
%in practice \newt{however, one may not be in this regime.} 

We will construct effective Hamiltonians using the two-body reduced density matrix (2-RDM) elements, 
\newt{given by the estimator (the normalization has been omitted),}
\begin{widetext}
\begin{eqnarray}
	\rho_{ijkl} &\equiv& \langle c_{i}^{\dagger} c_{j}^{\dagger} c_l c_k \rangle = \sum_{ a \neq b} \int \phi^{*}_k (r'_a) \phi^{*}_l (r'_b) \phi_i(r_a) \phi_j(r_b) \Psi^{*}({\bf{R}''_{ab}})~\Psi ({\bf{R}}) dr'_a dr'_b d {\bf{R}},
\label{eq:2rdm}
\end{eqnarray}
\end{widetext}
where $ {\bf{R}''_{ab}} = (r_1,r_2, r'_a..., r'_b,....r_N)$ refers to the set of 
\newt{electron coordinates obtained by changing the location of two electrons
and $\phi_i(r)$ are a chosen set of} one-particle wavefunctions (orbitals) indexed by 
label $i (j,k,l)$. The mixed estimator equivalent of Eq.~\eqref{eq:2rdm} 
is obtained like that for the energy. 
More details of this computation 
have been previously discussed elsewhere by one of us~\cite{Wagner_JCP}. The chosen set of orbitals 
is often localized on the atoms; this property 
makes it convenient to derive Hubbard-like models. 
We explain their construction next. 

\subsection{Localized orbitals}
\label{subsec:wannier}
Localized orbitals often provide an intuitive way of understanding an electronic 
system in terms of electron hops and on-site or inter-site repulsions. 
Thus, many works have been devoted to this subject; ranging from the Linearized Muffin-Tin 
Orbital (LMTO) method~\cite{Andersen} to the maximally localized Wannier 
function construction~\cite{Marzari_Wf}. 
Orbital localization has also been widely discussed 
in the quantum chemistry literature. 

\begin{figure}[htpb]
\centering
\includegraphics[width=\linewidth]{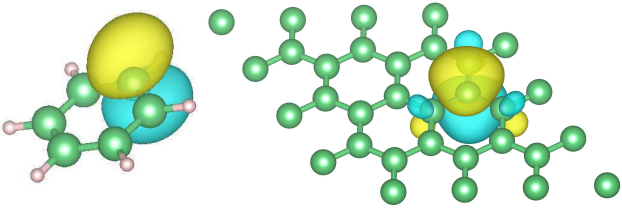}
\caption{\newt{Amplitude isosurface contour for (left) one of the six symmetry equivalent $\pi$ orbitals of the 
benzene molecule, obtained by localizing three bonding 
and three anti-bonding orbitals and (right) a representative localized 
orbital for the $4\times4$ unit cell of graphene. The colors indicate the sign 
of the single particle wavefunction.}}
\label{fig:orbs} 
\end{figure}	

The idea is to first select a set of orbitals 
in a certain energy window. For solids these correspond to 
bands close to the Fermi level, and for molecules 
these are partially occupied orbitals which constitute the active space. 
Then, a unitary transformation is performed to optimize a 
pre-decided metric for localization. 
In this work, we minimize the spread $S$,
\begin{equation}
S \equiv \sum_{n} \left( \langle \phi_n | r^2 | \phi_n \rangle - \langle \phi_n | {\bf{r}} | \phi_n \rangle^{2} \right)
\end{equation} 
where $\phi_n(r)$ are the desired localized orbitals related to the chosen set of 
orbitals $\Phi_i(r)$ by a unitary transformation, 
$\phi_n (r) = \sum_{i} U_{ni} \Phi_i(r)$.  

For some systems, as we will see in the case of benzene and graphene, 
schematically shown in Fig.~\ref{fig:orbs}, it is necessary to include unoccupied orbitals 
to get well-localized orbitals of the right symmetry~\cite{Hansen}. 
Thus the construction of localized orbitals is not a black-box procedure and 
may need adaptations based on the specifics of the system.

\newt{We note that the optimized parameters of the effective Hamiltonian may, 
in general, depend on the choice of localized orbitals. 
However, we have not explored this dependence - our main objective in this paper 
is to assess the validity of model Hamiltonians with respect to 
{\it ab initio} calculations for a particular choice of single particle basis.}  
 
\subsection{Lattice model calculations}
The lattice model calculations for Hubbard models of 
benzene and graphene at half-filling were carried out with a combination of our own codes 
and the freely available QUEST determinantal quantum Monte Carlo package~\cite{QUEST}. 
For the honeycomb lattice half-filled Hubbard model, the determinantal QMC 
method is sign problem free and the results are exact up to statistical errors.
A time step of $0.1$ was chosen and $\beta$ (the imaginary time) was set to $20$ 
for every calculation. Measurements were performed for $5000$ sweeps, 
with an additional $2000$ sweeps being used for equilibration.

\section{Criteria for a low energy effective Hamiltonian}
\label{sec:demands}
Our aim is to obtain a low energy effective Hamiltonian defined in 
the active space of electrons.
In this basis, the criteria it must satisfy are:
\begin{enumerate}[(a)]
\item  The reduced density matrices (RDM) of the ground and excited states obtained 
          from the {\it ab initio} calculation must match with that of the model calculation. 
\item  The energy spectra of the {\it ab initio} and model systems must match in the energy window of interest. 
\item The model must be detailed enough to capture the essential physics 
         and yet simple enough to avoid over-parameterization and over-fitting. 
\end{enumerate}

The concept of matching RDMs [criterion (a)] 
has previously appeared in related contexts~\cite{Acioli,Zhou_Ceperley, Changlani_percolation} 
and in work by one of us~\cite{Wagner_JCP}. Most physical properties, 
such as the charge and spin structure factors, are functions of the 2-RDM.
Since it is computationally expensive to calculate high-order RDMs, 
we use the matching condition only on the 2-RDM, 
$\rho_{ijkl} \equiv \langle c_i^{\dagger} c_j^{\dagger} c_l c_k \rangle$ 
where $i,j,k,l$ are orbital indices (including space and spin). 
This criterion automatically ensures that the combined number of 
electrons occupying the orbitals is equal to those in the model Hamiltonian. 
If any input state does not have the expected electron number 
in the active space, it can not be described by the effective Hamiltonian.  

The importance of excited state energies 
used in the fitting [criterion (b)] is highlighted by the fact that 
the wavefunctions, and their corresponding two-body density matrices, 
are invariant to many kinds of terms that enter the Hamiltonian.
For example, the transformation, 
\begin{equation}
	H' \rightarrow H + \alpha S^2 + \beta n + \gamma S^2 n 
\end{equation}
is, by construction, consistent with all the 2-RDM data 
for any $\alpha$, $\beta$, $\gamma$ for systems which have definite spin symmetry 
and particle number. Imposing certain physical 
constraints on the form of the interactions can reduce 
the need for this criterion. To give a concrete example, 
consider wavefunction data generated from the ground state of an unfrustrated 
Heisenberg spin Hamiltonian \newt{on} a bipartite lattice
~\cite{Note1}, $ H = J \sum_{\langle i,j \rangle}\bar{S}_i \cdot \bar{S_j}.$ 
where $\langle i,j \rangle$ refer to nearest neighbor pairs \newt{
and $\bar{S}_i$ is the spin operator on site $i$.}
Then adding $\alpha S^2$ gives the same reduced density matrices 
for the ground state, as long as $\alpha$ is small enough to not cause 
energy crossings i.e. not make an original 
excited state the new ground state. This additional 
term has the effect of introducing long-range Heisenberg couplings. 
Moreover, the effective Hamiltonian is not unique; 
the Lieb-Mattis model~\cite{LM} $H = S_A \cdot S_B$ 
(where $A$($B$) and $S_{A(B)}$ refer to the sublattice and corresponding spin), 
is also known to reproduce the low-energy limit of the Heisenberg model. 
Thus, imposing the requirement that the Hamiltonian has the 
nearest-neighbor form constrains $\alpha$ to zero and picks 
one particular model. Similar arguments should 
apply to extended Hubbard models in homogeneous 
systems where a physical constraint is that the 
density-density interaction must decrease monotonically with distance between orbitals.

\section{ {\it Ab Initio} Density matrix based Downfolding (AIDMD) procedures}
\label{sec:fitting}

We now discuss two procedures that use density matrices and 
energies to calculate parameters entering 
the effective Hamiltonian; both have been 
schematically depicted in Fig.~\ref{fig:hamfit}. 

\subsection{Eigenstate AIDMD method}
 
In the first method, schematically depicted in Fig.~\ref{fig:hamfit}(a), 
eigenstates from an {\it ab initio} calculation
are used to match density matrices and energies of the corresponding model. 
The QMC extrapolated estimator is taken to be an accurate representation of the true one.
Then the parameters of the model Hamiltonian are obtained by  
minimizing a cost function that is a linear combination 
of the energy and density matrix errors,
\begin{equation}
	\mathcal{N} \equiv \sum_{s} (E_s^{a}-E_s^{m})^{2} + f \sum_{s} \sum_{i,j,k,l} (\langle c_i^{\dagger} c_j^{\dagger} c_l c_k \rangle^{a}_{s} - \langle c_i^{\dagger} c_j^{\dagger} c_l c_k \rangle^{m}_{s} )^2 
\end{equation}
where the subscript $s$ is an eigenstate index, $i,j,k,l$ are orbital indices 
and the superscripts $a$ and $m$ refer to {\it ab initio} and model 
calculations respectively. There is no definite prescription for choosing 
the weight $f$;
a good heuristic is to choose a value that gives roughly the same size 
of errors for the two terms that enter the cost function. 
The cost minimization is performed with the Nelder Mead simplex algorithm. 

In practice we found that since the number of 
available eigenstates and the accuracy of true estimators 
is limited, a second method discussed next 
is more suited for downfolding. 

\subsection{Non-eigenstate AIDMD method}
\label{sec:AxE}
Consider a set of {\it ab initio} energy averages $\tilde{E}_s$, i.e. expectation values of the Hamiltonian, 
and corresponding 1- and 2-RDMs $\langle c_i^{\dagger} c_j \rangle_s$, 
$\langle c_i^{\dagger}c_j^{\dagger} c_l c_k \rangle_s$ 
for arbitrary low-energy states characterized by index $s$. 
Assume a model 2-body Hamiltonian with 
effective parameters $t_{ij}$ (1-body part) 
and $V_{i,j,k,l}$ (2-body part) along with a constant term $C$; 
the total number of parameters being $N_p$. 
Then for each state $s$, we have the equation, 
\begin{equation}
	\tilde{E}_s \equiv \langle H \rangle_s = C + \sum_{ij} t_{ij} \langle c_i^{\dagger} c_j \rangle + \sum_{ijkl} V_{ijkl} \langle c_i^{\dagger}c_j^{\dagger} c_l c_k \rangle  
\end{equation}
where we have made the assumption that the chosen set of operators 
corresponding to single particle wavefunctions or orbitals, 
explain all energy differences seen in the 
{\it ab initio} data. The constant $C$ is from energetic contributions 
of all other orbitals which are not part of the chosen set.  

We then perform calculations for $M$ low-energy states 
which are not necessarily eigenstates. \newt{ These states are not arbitrary 
in the sense that they have similar descriptions of the core and virtual spaces. 
Each state satisfies the criteria (1) its energy average 
does not lie outside the energy window of interest and (2) the trace of its 
1-RDM matches the electron number expected in the effective Hamiltonian.} 

\newt{The objective of choosing a sufficiently big set of states is to explore 
parts of the low-energy Hilbert space that show variations 
in the RDM elements.} Since the same parameters describe all $M$ states, 
they must satisfy the linear set of equations, 
\begin{align}
\left(
\begin{array}{c}
\tilde{E}_1 \\
\tilde{E}_2 \\
\tilde{E}_3 \\
... \\
... \\
... \\
... \\
\tilde{E}_M
\end{array}
\right) =
\left(
\begin{array}{ccccc}
1 & \langle c_i^{\dagger}c_j \rangle_{1}  & .. & \langle c_i^{\dagger}c_j^{\dagger}c_l c_k \rangle_{1} & .. \\
1 & \langle c_i^{\dagger}c_j \rangle_{2}  & .. & \langle c_i^{\dagger}c_j^{\dagger}c_l c_k \rangle_{2} & .. \\
1 & \langle c_i^{\dagger}c_j \rangle_{3}  & .. & \langle c_i^{\dagger}c_j^{\dagger}c_l c_k \rangle_{3} & .. \\
1 & \langle c_i^{\dagger}c_j \rangle_{4}  & .. & \langle c_i^{\dagger}c_j^{\dagger}c_l c_k \rangle_{4} & .. \\
1 & ....                                  & .. & ..                                                    & .. \\
1 & ....                                  & .. & ..                                                    & .. \\
1 & \langle c_i^{\dagger}c_j \rangle_{M}  & .. & \langle c_i^{\dagger}c_j^{\dagger}c_l c_k \rangle_{M} & .. \\
\end{array}
\right) \left(
\begin{array}{c}
C           \\
t_{ij}      \\
..          \\
V_{ijkl}    \\
..
\end{array}
\right)
\label{eq:bigE_Ax}
\end{align}
which is compactly written as,
\begin{equation}
	{\bf{E}} = A {\bf{x}}
\label{eq:E_Ax}
\end{equation}
where $ {\bf{E}} \equiv (\tilde{E}_1,\tilde{E}_2,...\tilde{E}_M)^{T}$ 
is the $M$ dimensional vector of energies, $A$ is the $M \times N_p$ matrix composed 
of density matrix elements and $ {\bf{x}} \equiv (C,t_{ij}....V_{ijkl}...)^T$ 
is a $N_p$ dimensional vector of parameters.
This problem is over-determined for $M>N_p$, which is the regime we expect to work in.

\begin{figure}[htpb]
\centering
\includegraphics[width=1\linewidth]{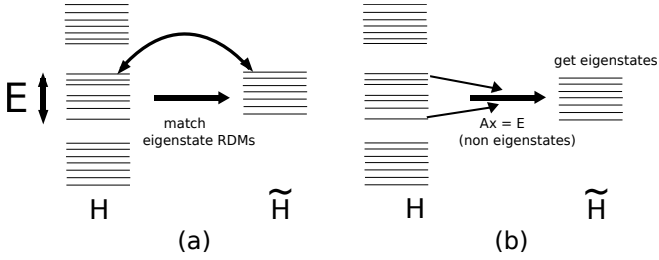}
\caption{Schematic of {\it ab initio} density matrix downfolding (AIDMD) 
methods employed for determining the effective Hamiltonian parameters. 
(a) In the eigenstate (E)-AIDMD, the reduced density matrices and energies 
of eigenstates of the model \newt{are} matched to the {\it ab initio} 
counterparts. (b) The non-eigenstate (N)-AIDMD method uses 
RDMs and energies of arbitrary low-energy states to construct a matrix of relevant 
density matrices and performs a least square fit to determine the optimal parameters. }
\label{fig:hamfit} 
\end{figure}	

In the case of any imperfection in the model, which is the most common case, 
the equality will not hold exactly 
and one must then instead minimize the norm of the error, $\mathcal{N}$:
\begin{equation}
	\mathcal{N} \equiv ||A\bf{x}-\bf{E}||^2
\label{eq:norm}
\end{equation}
This cost function can be minimized in a single step by using 
the method of least squares, employing the singular 
value decomposition of matrix $A$. This matrix also encodes exact (or near-exact) linear dependences. 
Thus, the quality of the fit is directly judged 
by assessing (1) the singular values of the $A$ matrix and (2) 
the value of the cost function itself i.e. the deviations of the input and fitted energies.
We will refer to this as the non eigenstate (N)-AIDMD method throughout the rest of the paper.
This idea is schematically depicted in Fig.~\ref{fig:hamfit}(b).

The matrix $A$ gives a very natural basis for understanding renormalization effects.
For example, consider a set of wavefunctions that show that the correlator 
$\langle n_i n_j \rangle$ does not change significantly. This would lead to the 
corresponding column of matrix $A$ being identical (up to a scale factor) 
to the first column of 1's. Physically, this would correspond to the coupling constant 
$V_{ijji}$ being irrelevant for the low-energy physics; 
it can take any value including zero 
and can be absorbed into the constant shift term. 
This could also alternatively mean that the input 
data is correlated and does not provide enough information about $V_{ijji}$, so care must be taken in constructing the set of wavefunctions.

In summary, the N-AIDMD method performs the following operation. 
The \newt{1- and} 2-RDMs and energy expectation values of many non-eigenstate correlated states are calculated.
Then, given an effective Hamiltonian parameterization, linear equations~\eqref{eq:E_Ax} are 
constructed and solved. Standard model fitting principles apply, and we can 
evaluate the goodness of fit to determine whether a given effective Hamiltonian 
can sufficiently describe the data.

\subsection{Generating states for AIDMD methods}
\label{sec:state_generation}

\newt{We now address the central issue of generating states to be used as inputs for the AIDMD methods.

For the E-AIDMD, the near-eigenstates were obtained by 
performing CISD calculations with multiple roots and optimizing 
a multi-determinant Jastrow wavefunction with each CISD guess as a starting point. 
This is known to be approximate, especially for higher excited states. It is the 
inherent uncertainty about the accuracy of this procedure, along with the 
fact that only a small number of eigenstates are accessible, that limits the utility of E-AIDMD. 
 
From the point of view of the N-AIDMD method too, automating the construction of the 
database of wavefunctions may not be completely straightforward and here we offer 
some heuristics for doing this within QMC methods. We re-emphasize that any state described by 
the effective Hamiltonian must be one that does not involve large contributions 
from the core and virtual orbitals i.e. single particle degrees of freedom outside 
the chosen active space. This check can be imposed at the {\it ab initio} level by monitoring the 
RDMs, for example, the trace of the 1-RDM taken over the 
active orbitals must equal the number of electrons in the effective Hamiltonian description. 

One way to generate new states is to perturb near-eigenstates. For example, 
after optimizing the multi-determinantal-Jastrow trial wavefunction, we artificially change the 
determinantal coefficients by small amounts. This procedure changes the nodal surface and gives 
energies close to, but different from, the optimized ground state. A second source of data is spin excitations of the DFT 
reference Slater determinant, generated within the space of orbitals that play an important role 
in the active space; for benzene and graphene these involve the Kohn-Sham orbitals with 
$\pi$ symmetry. Finally, in the case of extended systems, we chose a linear combination of determinants which, 
in spite of being not size-extensive, reveal additional properties of the effective Hamiltonian.} 

\subsection{Quantum Monte Carlo specific adaptation}
\label{sec:qmc_param}
The formalism introduced above applies to any method 
that calculates energies and density matrices. 
In this paper, all the expectation values entering the $A$ matrix 
are calculated for the chosen low-energy wavefunctions by Monte Carlo sampling.

\newt{Once this database of states has been generated, we perform 
two independent calculations to estimate the parameters, 
one using variational and the other using mixed estimators.} 
In the latter case, we modify the linear equations~\eqref{eq:E_Ax} by using 
$E_s = \langle {\psi_T}^{s}| H | {\psi}^s \rangle$ and 
the projected estimates of the 
density matrix elements i.e. $\langle {\psi_T}^s | c_i^{\dagger} c_j | {\psi}^s \rangle$ 
and $\langle {\psi_T}^s| c_i^{\dagger}c_j^{\dagger} c_l c_k | {\psi}^s \rangle$ 
in the construction of the $A$ matrix. 
The implicit normalization of these mixed estimates by 
$\langle {\psi_T}^s | \psi^{s} \rangle$ is assumed. 
This projector formulation 
is also amenable to coupled-cluster calculations which 
work with projected energies and density matrices.

The bias arising out of fixing the nodes of the projected wavefunction 
does not affect our formulation. This is because the method regards $\psi \psi_T$ 
as some arbitrary positive sampling function associated with a low energy state, 
and it is this same distribution that is used for 
the evaluation of the density matrix elements. 
Each such distribution provides a linear equation encoding the 
relationship between the FN-DMC energy and the projected density matrix elements 
and the unknown parameters. Up to errors from statistical uncertainties 
and from the assumption of the form of the Hamiltonian, this relationship is exact. 

\newt{However, the value of the optimal parameters does depend on the choice of method.} 
For example, for the benzene molecule presented in section~\ref{sec:benzene}, 
the VMC and FN-DMC parameter values agree with each other up 
to $10\%$; the largest discrepancy is due to 
different constant terms. 
This discrepancy is expected, because VMC does not 
provide an accurate description of the core and virtual spaces.  

Ideally, only the FN-DMC calculations should be used 
to estimate the parameters. However, the mixed estimator 
in FN-DMC is biased because of the inaccuracy of the 
trial wavefunction. Thus, we propose a better 
estimator for the parameters, 
\begin{equation}
	p = 2 p_{D} - p_{V}
\end{equation}
where $p$ is the true parameter, 
and $p_D$ and $p_V$ are the corresponding parameters 
obtained from FN-DMC and VMC calculations. 
The details of this result are explained in Appendix A. 

\section{Simple application: Hubbard to Heisenberg model}
To demonstrate our formalism for a simple example, we 
consider the two site Hubbard model and fit information from the lowest two states 
to a Heisenberg model. 
%This is not technically downfolding, as there is no reduction 
%of electron number. 

We analytically solve for all four eigenstates of the Hamiltonian,
\begin{equation}
H = -t \sum_{ij} c_{i,\sigma}^{\dagger} c_{j,\sigma} + \text{h.c.} + U \sum_{i} n_{i,\uparrow} n_{i,\downarrow}
\label{eq:two_hubbard}
\end{equation}
for two opposite spin electrons on two sites, where $t$ is the hopping, $U$ is the Hubbard on-site interaction.
The Hilbert space on a single site (orbital) is spanned by four states $|0 \rangle$ (unoccupied), $|\uparrow \rangle$ (single up occupied), $|\downarrow \rangle$ (single down occupied), $|\uparrow\downarrow \rangle$ (doubly occupied). 
For completeness, we discuss some features of the solution method below.
 
First notice that the triplet state $| \psi_t \rangle \equiv \frac{|\uparrow \;\; \downarrow \rangle - |\downarrow \;\; \uparrow \rangle }{\sqrt{2}} $ with energy $E_t=0$ and the state 
$|\psi_d \rangle \equiv \frac{| \uparrow \downarrow \;\; 0 \rangle - | 0 \;\; \uparrow\downarrow \rangle }{\sqrt{2}}$ 
with energy $E_d=U$, are exact eigenstates of the problem independent of the values of $t$ and $U$. 

To get the other two states, write the Hamiltonian in the 
basis of $ |\psi_s \rangle \equiv \frac{1}{\sqrt{2}} \left( |\uparrow \;\; \downarrow \rangle + |\downarrow \;\; \uparrow \rangle \right) $ 
and $ | {\psi_d}^{'} \rangle \equiv \frac{1}{\sqrt{2}} \left( |\uparrow\downarrow 0 \rangle + |0 \uparrow\downarrow \rangle \right) $, 
\begin{eqnarray}
H =
\left(
\begin{array}{cc}
0   & -2t \\
-2t & U    \\
\end{array}
\right)
\end{eqnarray}
Then diagonalizing it, we get the energies to be, 
\begin{equation}
E_{\pm} = \frac{U \pm \sqrt{U^2+16 t^2} }{2}
\end{equation}

The lowest energy corresponds to the singlet, $E_{-}$ and the 
corresponding eigenvector is 
\begin{equation}
	| \psi_{-} \rangle = \frac{2t}{\sqrt{4t^2+{E_{-}}^2}} |\psi_s \rangle  - \frac{E_{-}}{\sqrt{4t^2+{E_{-}}^2}} |{\psi_d}^{'} \rangle
\end{equation}
with the next excited state being the triplet $|\psi_t \rangle$.

We choose the Heisenberg form to fit to
\begin{equation}
	\tilde{H} = C + J S_1 \cdot S_2
\end{equation}
To determine the parameters $C$ and $J$, form the $2 \times 2$ $A$ matrix 
with the lowest two energy states,
\begin{equation}
\left(
\begin{array}{c}
 E_{-} \\ 
 E_t \\
\end{array}
\right)
=
\left(
\begin{array}{cc}
1   &  \frac{1} {4+(E_{-}/t)^2}	\\
1   & -3/4                      \\
\end{array}
\right)
\left(
\begin{array}{c}
 C \\
 J \\
\end{array}
\right)
\end{equation}
Using derived values of $E_s$ and $E_t=0$, we get,
\begin{equation}
	J = \frac{E_s (4+(E_{-}/t)^2)}{4+\frac{3}{4}(E_{-}/t)^2}
\end{equation} 
which to lowest order in $t/U$ is $J = -4t^2/U$.

Observe that the correlator for $\langle S_i \cdot S_j \rangle$ is not exactly 1/4 
but only approximately so. This is expected since the fluctuations from the 
high-energy states are not exactly zero, if it were, it would be equivalent to 
exactly block-diagonalizing the Hamiltonian. This exact 
block diagonalization is not possible in general, unless it is 
also accompanied with a change in the low energy degrees of freedom 
entering the model.

If we now rotate the two low energy eigenstates and define the orthogonal linear combinations,
\begin{subequations}
\begin{eqnarray}
	| \psi_1^{'}\rangle &=& p | \psi_{-} \rangle + q | \psi_t \rangle \\ 
	| \psi_2^{'}\rangle &=& q | \psi_{-} \rangle - p | \psi_t \rangle 
\end{eqnarray}
\end{subequations}
and calculate their energies and construct the corresponding $A$ matrix, 
we get $J$ to be independent of $p$ and $q$. This property is 
desirable as the method does not hinge on the requirement of 
eigenstates as inputs.

In terms of canonical transformations (here equivalent to second order perturbation theory), 
the matrix element of the effective Hamiltonian between the singly occupied states is, 
\begin{eqnarray}
	\langle \uparrow\downarrow | \tilde{H} | \downarrow \uparrow \rangle &=& \sum_{n} \frac{ \langle \uparrow\downarrow|H| n\rangle\langle n |H| \downarrow \uparrow \rangle } {0-E_n} \nonumber \\ 
	                                                                     &=& \frac{-2 t^{2}}{U}
\end{eqnarray}
Since this matrix element must equal $J/2$ in the Heisenberg model we arrive at the same result, 
namely $J = -4t^{2}/U $~\cite{Note2}.
%\footnote{To map a fermion model to a spin model, a 
%Jordan-Wigner phase is introduced which remains fixed, 
%as there are no charge fluctuations. 
%In this notation the relative signs between basis states 
%entering in the singlet and triplet states is reversed 
%and $J=+4t^2/U$}.

%To map a fermion model to a spin model, a 
%Jordan-Wigner phase is often introduced which is fixed as there are no charge fluctuations. 
%In this notation the singlet is $ \frac{1}{\sqrt{2}} ( |\uparrow \downarrow \rangle - |\downarrow \uparrow \rangle ) $, 
%the triplet is $ \frac{1}{\sqrt{2}} ( |\uparrow \downarrow \rangle + |\downarrow \uparrow \rangle ) $ 
%and $J=+4t^2/U$.

\section{Application to a molecule: Benzene}
\label{sec:benzene}
We show the workings of the described AIDMD methods for the benzene molecule. 
Our choice of system is motivated by its simplicity 
(one band model and presence of many symmetries) 
and the availability of experimental energies to compare to. 

DFT calculations were first performed in the TZVP basis with Burkatzki-Filippi-Dolg pseudopotentials~\cite{BFD}, 
using the molecular geometry that corresponds to a B3LYP optimized calculation. 
These serve as a starting point for the QMC calculations, discussed in the Methods section.
In the charge neutral sector, there are a total of 30 electrons 
(\newt{for example,} 15 $\uparrow$  and 15 $\downarrow$ for the spin-singlet state) 
and our objective is to downfold this system to an effective one with 6 electrons 
(3$\uparrow$, 3 $\downarrow$). 

The model Hamiltonian is defined in the space of six $\pi$ orbitals; 
a representative localized orbital has been shown in Fig.~\ref{fig:orbs}.
%These orbitals are not the atomic $p_z$ ones, but rather their renormalized versions.
These orbitals were obtained by localizing 
the highest three occupied and the lowest three 
unoccupied B3LYP orbitals (from the $S=0$ DFT calculation) 
with $\pi$ orbital symmetry, a well established procedure 
in the literature~\cite{Hansen}. The overall phase of these orbitals is 
adjusted to enable use of parameter symmetries directly when fitting.

Appendix B discusses the QMC data used for fitting and 
the initial pre-processing to determine the eligibility 
of states that can be described by a six-$\pi$ orbital Hamiltonian. 

\subsection {On-site Hubbard model}
We consider the Hubbard model for six orbitals of benzene, 
given by Eq.~\eqref{eq:two_hubbard}, where $t$ is used for 
the nearest-neighbor hopping and $U^{*}$ is 
the effective on-site Coulomb repulsion. We will discuss multiple ways 
of using reduced density matrix elements to estimate these parameters.

\subsubsection{Hubbard $U^{*}/t$ from the E-AIDMD method}
An estimate of $U^{*}/t$ is obtained 
by directly matching the half-filled ground state ($S=0$) 2-RDM element corresponding 
to the "double occupancy" correlator ($\langle n_{\uparrow} n_{\downarrow} \rangle$) 
of the {\it ab initio} and lattice-model calculations. This element equals $0.25$ 
for the non interacting case ($U^{*}=0$) and its value reduces for $U^{*}>0$.

\begin{figure}[htpb]
\centering
\includegraphics[width=1\linewidth]{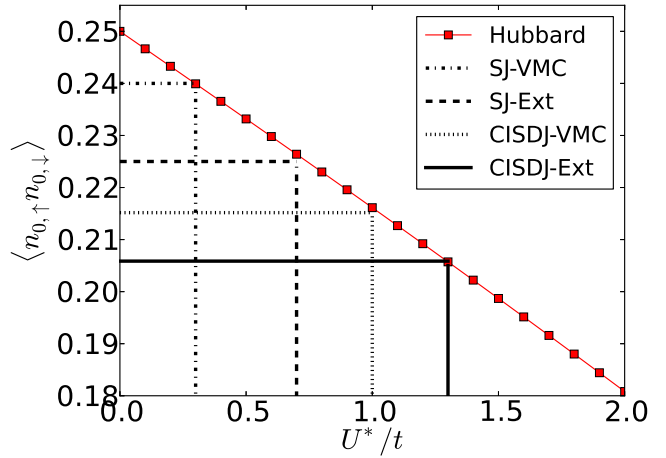}
\caption{Double occupancy correlator of a single $\pi$ orbital of a 
six-site ring Hubbard model, as a function of $U^{*}/t$, computed 
in the half-filled singlet ground state. 
Comparisons are made with the values from the variational (VMC) 
and extrapolated (Ext) estimators obtained from {\it ab initio} 
QMC calculations for the benzene molecule with optimized Slater-Jastrow (SJ) and 
configuration interaction singles doubles-Jastrow (CISDJ) wavefunctions.}
\label{fig:rdm_benzene} 
\end{figure}	

\begin{figure*}[htpb]
\centering
\subfigure[]{\includegraphics[width=0.34\linewidth]{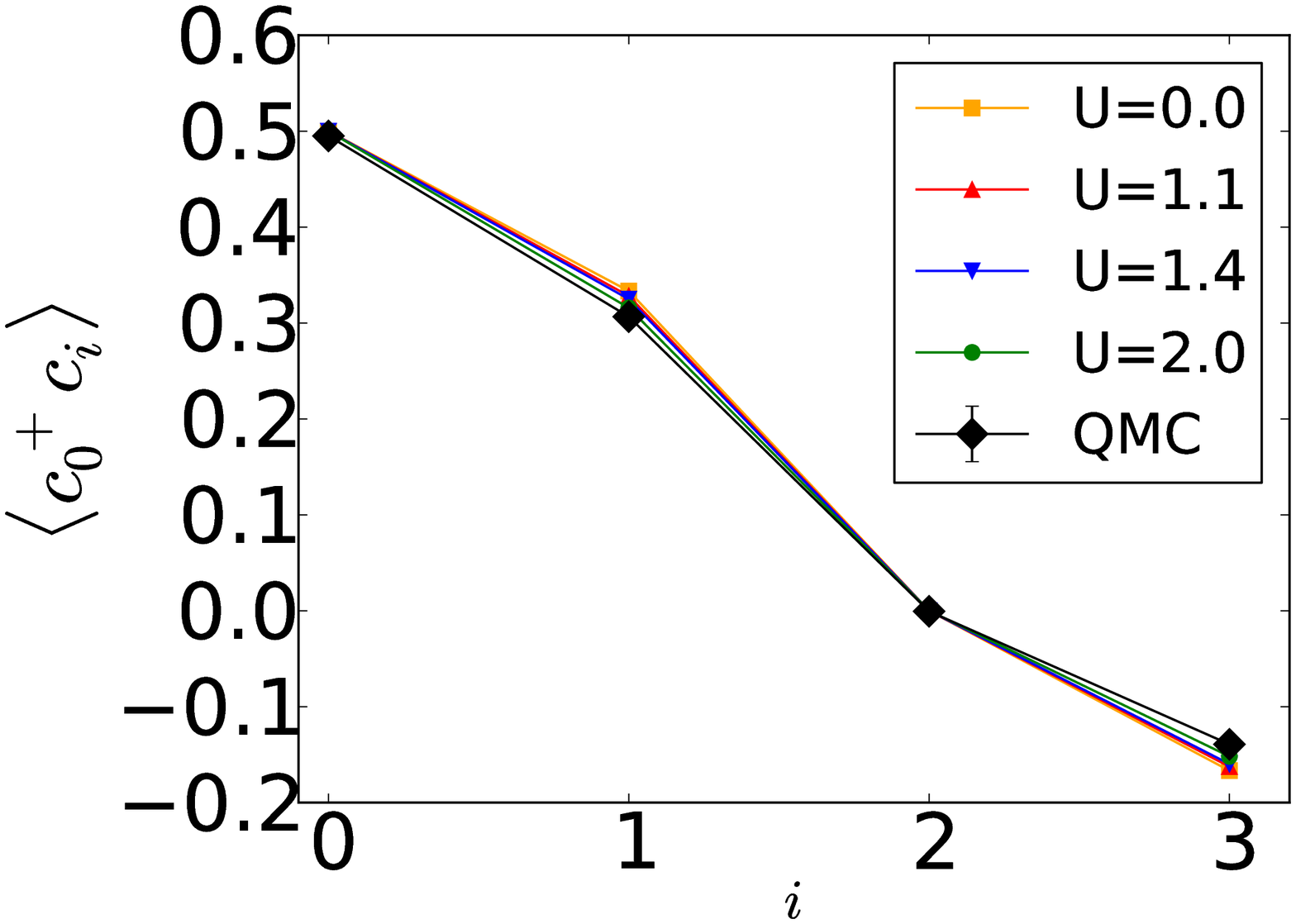}}
\subfigure[]{\includegraphics[width=0.32\linewidth]{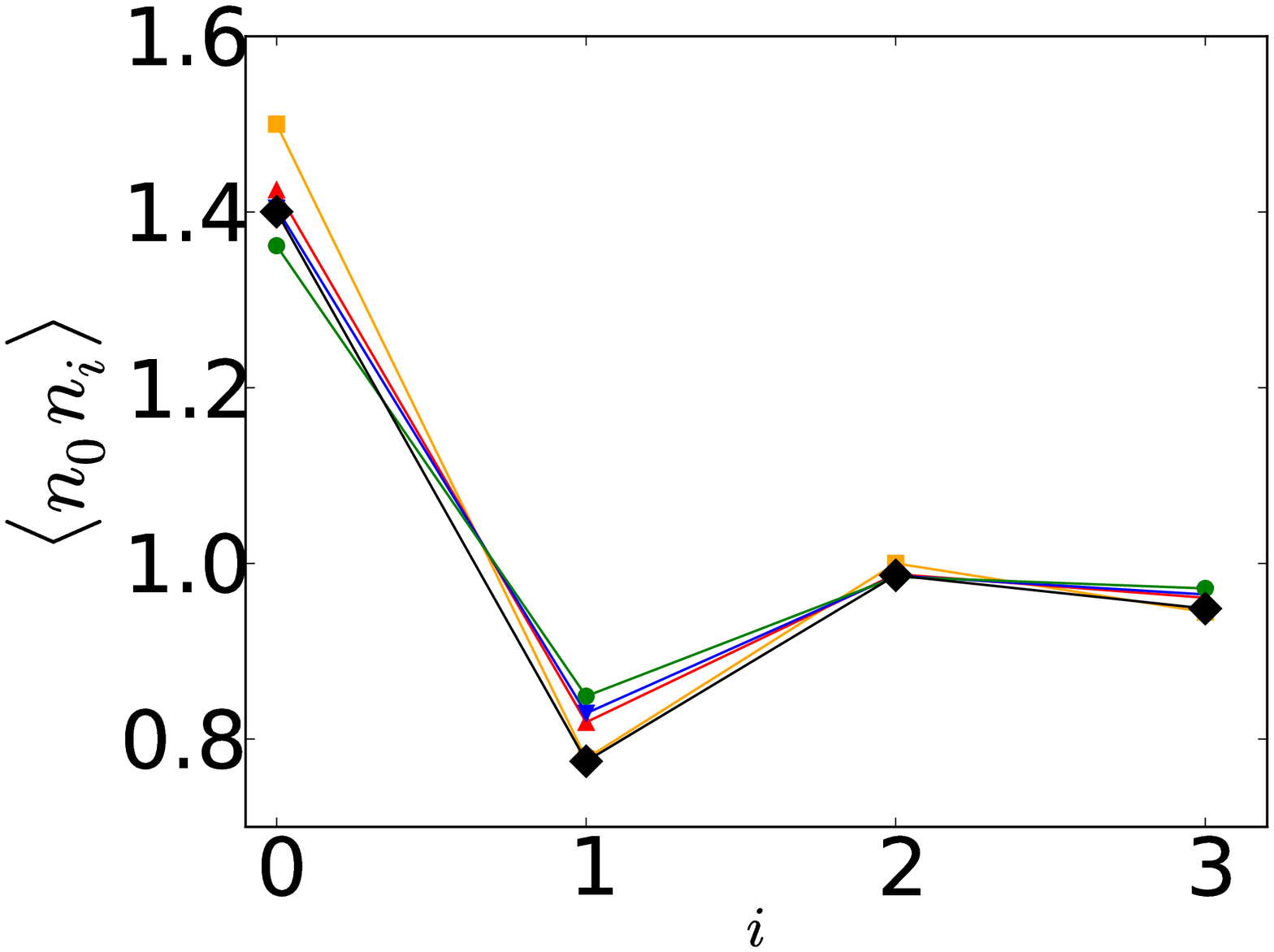}}
\subfigure[]{\includegraphics[width=0.32\linewidth]{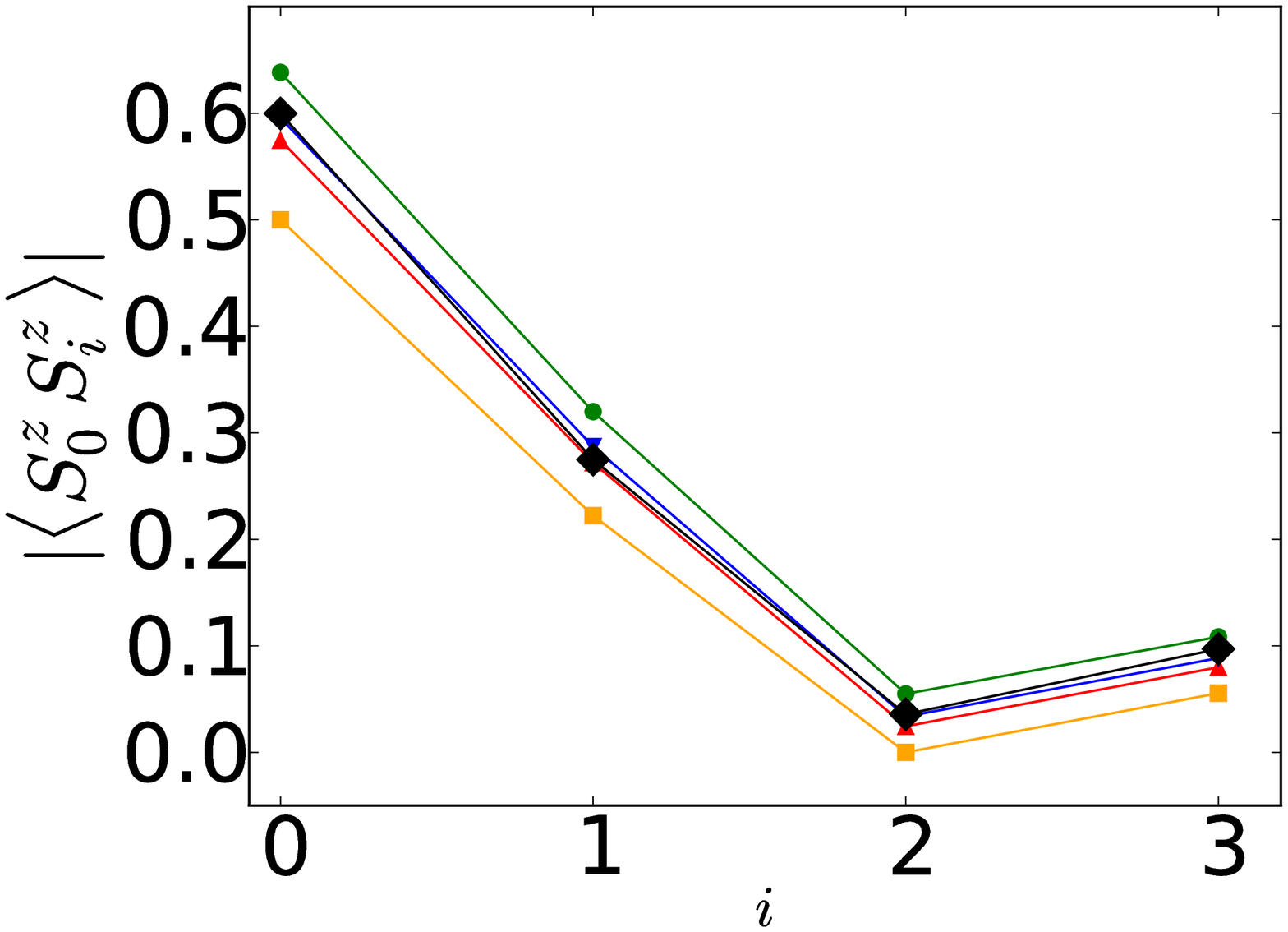}}
\subfigure[]{\includegraphics[width=0.32\linewidth]{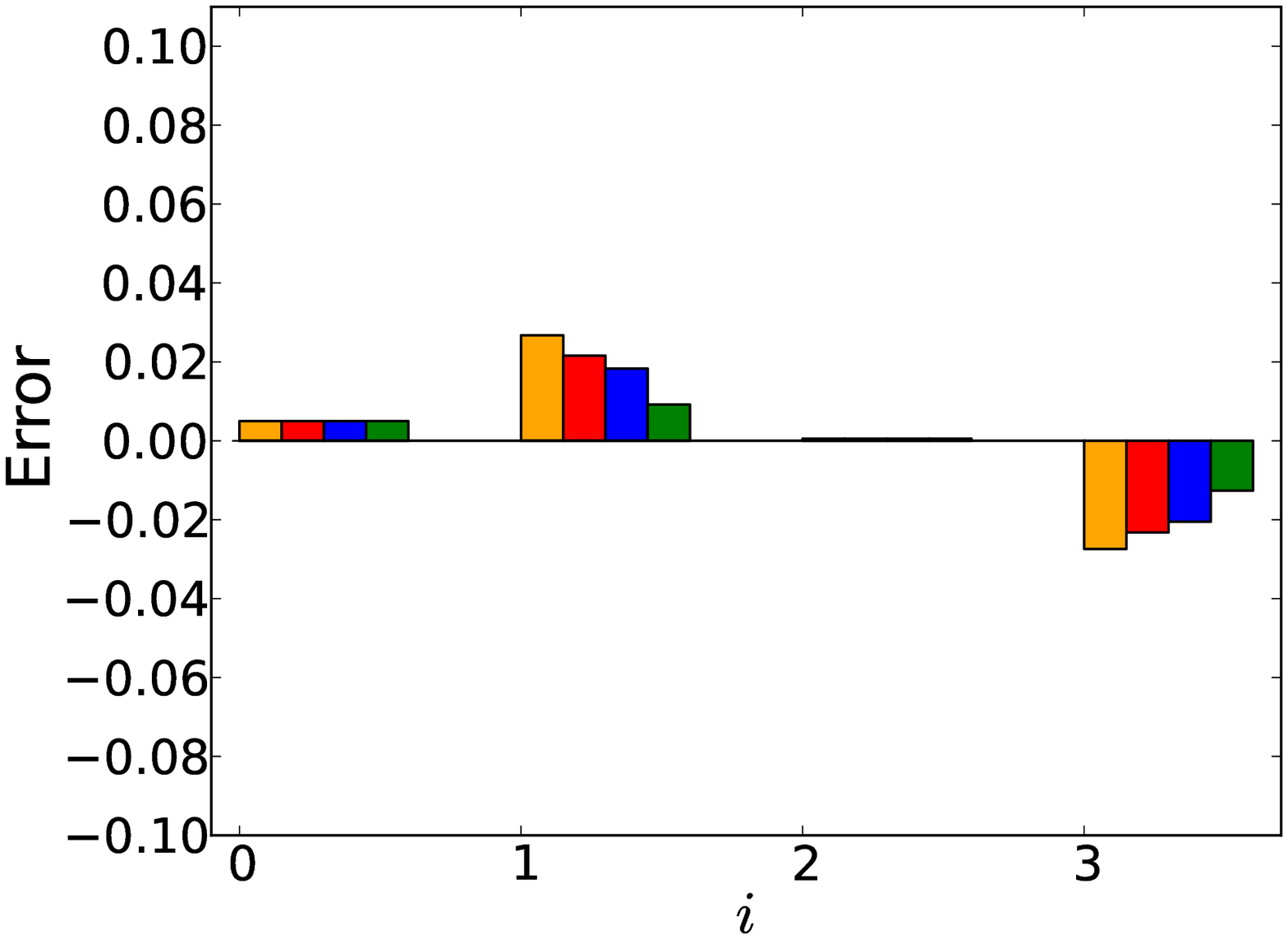}}
\subfigure[]{\includegraphics[width=0.32\linewidth]{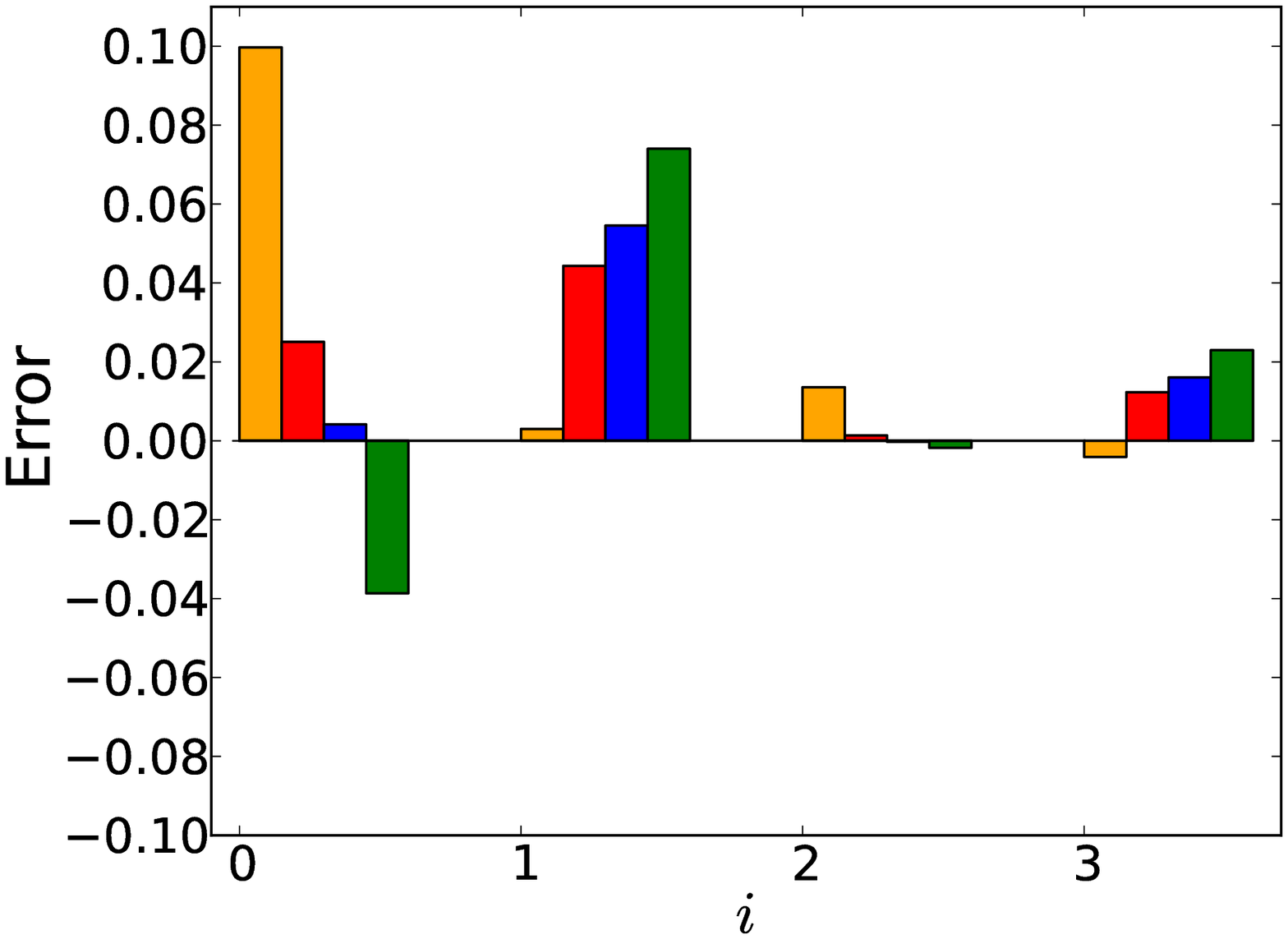}}
\subfigure[]{\includegraphics[width=0.32\linewidth]{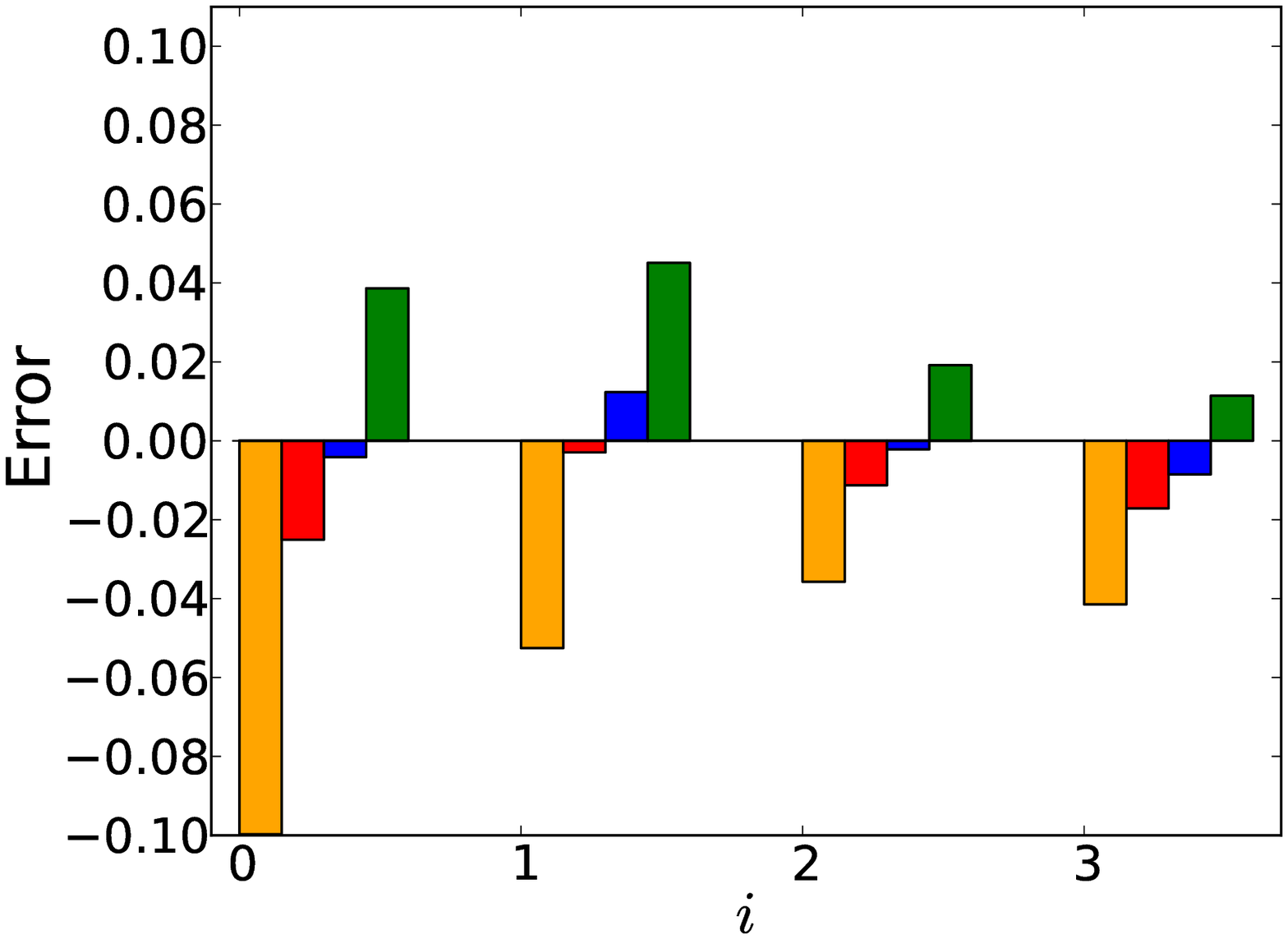}}
\caption{ \newt{Comparison of correlation functions from 
{\it ab initio} quantum Monte Carlo calculations of the benzene molecule 
and the six-site Hubbard ring for varying values of $U^{*}/t$. 
The half-filled ground state has been considered which corresponds to 15 $\uparrow$ 
and 15 $\downarrow$ electrons in the {\it ab initio} calculation and 3 $\uparrow$ and 
3 $\downarrow$ electrons in the model calculation. The top panels show the (a) one-body 
density matrix $\langle c^{\dagger}_{0,\sigma} c_{i,\sigma} \rangle$ (b) 
density-density correlators $\langle n_0 n_i \rangle$ (c) spin-spin correlators $\langle S_0^{z} S_i^{z} \rangle$, 
all as a function of distance ($i$) with respect to a reference site ($0$). 
Panels (d)-(f) show the same data, but from the point of view of errors of the corresponding model correlation functions with respect 
to the {\it ab initio} data. The one-body density matrix shows relatively small errors
for all $U^{*}/t$, but the dependence on $U^{*}/t$ is more pronounced for the density-density and 
spin-spin correlators. We infer that the value $U^{*}/t \approx 1.4$ reproduces most correlators well, 
except the nearest-neighbor density-density correlator.} } 
\label{fig:HubbardU_benzene}
\end{figure*}	

Fig.~\ref{fig:rdm_benzene} shows the dependence of 
$\langle n_{\uparrow} n_{\downarrow} \rangle$, computed in the ground state 
of the Hubbard model of a six-site ring at half filling, on $U^{*}/t$. 
The plot also indicates the value of this correlator computed from various wavefunctions 
and estimators from {\it ab initio} QMC calculations of the benzene molecule. 
The trends are consistent with our expectations; 
the Slater-Jastrow (SJ) wavefunction at the VMC level underestimates 
the strength of the effective interactions, which is partly remedied by 
the extrapolated estimator from FN-DMC. However, the bias (systematic error) 
is expected to be large because of the considerable difference in the two estimates. 
This bias is reduced by the multi-determinantal-Jastrow (CISDJ) 
wavefunction we employed; the difference 
between the variational and extrapolated estimator is about 5\%. 

The value of $U^{*}/t$ is found to be extremely sensitive to the precise value of 
the double occupancy correlator, a change of a few percent (i.e. from $0.24$ to $0.20$) 
changes our estimate from $\approx 0.3$ to $1.3$ (i.e. a factor of almost $4$). 
In general, this observation suggests that it is crucial to look at various other 
elements of the 2-RDM and to look at alternate ways 
of estimating Hubbard parameters.

\newt{
In Fig.~\ref{fig:HubbardU_benzene} we compare results of other correlators from extrapolated 
QMC estimates with those from the on-site Hubbard model 
on a six site ring for varying values of $U^{*}/t$.
We focus on the one-body density matrix 
$\langle c^{\dagger}_{0,\sigma} c_{i,\sigma} \rangle$ (the values are the same for 
both spin indices $\sigma$), density-density correlators $\langle n_0 n_i \rangle$ and 
spin-spin correlators $\langle S_0^{z} S_i^{z} \rangle$, all as a function of distance ($i$) 
with respect to a reference site (labelled $0$). 
The value of $U^{*}/t \approx 1.4$ gives the smallest errors 
for most observables, except for the nearest-neighbor
density-density correlator which \newt{favors} a value of 
$U^{*}/t \sim 0$. In the limit that the model is perfect, all estimates must 
yield the same value; the differences reflect an inadequacy 
of the on-site Hubbard model in describing all the data. 
This is evidence for the need for long-range interactions.}

\subsubsection{Hubbard $U^{*}/t$ from the N-AIDMD method}
As mentioned previously, the idea of matching density matrix elements
is useful only for comparing exact eigenstates. However, it is 
difficult to construct eigenstates with very high accuracy in the {\it ab initio}
calculations and at times also for the equivalent model for large system sizes. 
This is why we appeal to the N-AIDMD method, 
introduced and explained in Sec.~\ref{sec:AxE}, 
which is relatively insensitive to the nature of the states input 
to the method. 
%effective Hamiltonian parameters. Studying different energy windows 
%provides a way of monitoring the frequency dependence of 
%the Hamiltonian parameters. 
For charge-neutral benzene, we construct the $A$ matrix 
by taking various states in a $10$ eV energy window above 
the ground state using VMC and DMC methods.

\begin{figure*}[htpb]
\centering
\includegraphics[width=\linewidth]{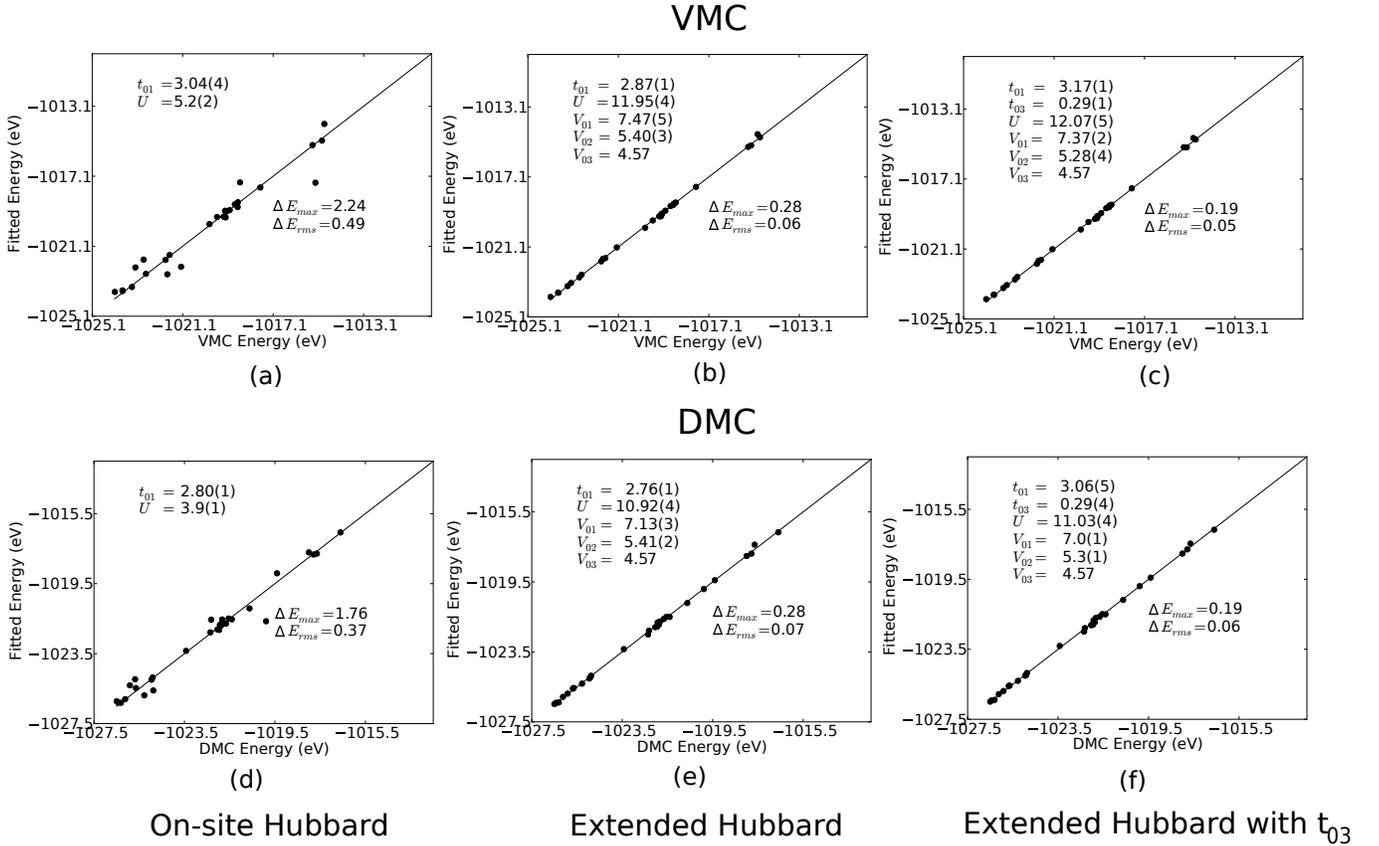}
\caption{\newt{Comparison of {\it ab initio} ($x$-axis) and fitted energies 
($y$-axis) of the benzene molecule using the N-AIDMD procedure for 
different model Hamiltonians. The non-eigenstate data is generated by considering 
various spin excitations and nodal surfaces, all in the same charge sector (30 electrons).
The {\it ab initio} energy is directly sampled using quantum Monte Carlo methods - 
for the variational Monte Carlo method (VMC) it corresponds to  
$\langle \psi_T | H | \psi_T \rangle / \langle \psi_T|\psi_T \rangle $ 
and for the fixed-node diffusion Monte Carlo 
(DMC) to $\langle \psi_T | H | \psi \rangle/ \langle \psi_T|\psi \rangle$ 
where $\psi_T$ and $\psi$ correspond to trial and projected wavefunctions 
respectively. 
The fitted energy is obtained from the optimized parameters by multiplying them 
by the corresponding {\it ab initio} density matrix elements. 
The top panels (a)-(c) show the VMC results and the bottom ones 
(d)-(f) show the fixed-node DMC ones. 
(a) and (d) correspond to the on-site Hubbard model 
on a six-site ring, (b) and (e) the extended Hubbard model, 
and (c) and (f) include an additional third-nearest neighbor hopping.}}
\label{fig:models} 
\end{figure*}	

Fig.~\ref{fig:models} shows the comparison 
of the fitted energy and the input VMC or DMC energy.  
The former is obtained by taking the linear combination of the 
{\it ab initio} VMC or DMC density matrices weighted by the optimized 
parameters \newt{of the effective Hamiltonian}. 
A perfect agreement between the fitted and \newt{{\it ab initio} } input data 
would correspond to all energies falling exactly on the $y=x$ 
line. By this measure, the Hubbard model for benzene 
is reasonable, though not accurate, as is seen 
in Fig.~\ref{fig:models}(a) \newt{and} (d). 
The presence of significant deviations of the order of $1-2$ eV from the $y=x$ 
line indicates the need for a more refined model, which we discuss in section~\ref{sec:exthubb}. 

The VMC data yields optimal parameters of $t=3.0$ eV and $U^{*}=5.2$ eV 
($U^{*}/t = 1.7$) and the FN-DMC data gives $t=2.8 $ eV and $U^{*}=3.9$ eV 
($U^{*}/t = 1.4$). The extrapolated estimate of the optimal parameters, 
$t=2.6$ eV agrees with the value of $t=2.54 $ eV reported by 
Bursill et al.~\cite{Bursill}. The extrapolated 
value of $U^{*}/t \approx 1.0$ is also broadly consistent
with a recently reported estimate~\cite{Schuler_graphene} to within 10-20\%.
  
%\subsubsection{Assessing the quality of the Hubbard model}
%
%Fig.~\ref{fig:HubbardU_benzene} shows comparison of the extrapolated 
%QMC estimates of the one-body density matrix, density-density and spin-spin correlators 
%and those for the Hubbard model for various values of $U^{*}/t$. 
%The value of $U^{*}/t = 1.4$ gives the smallest errors 
%for most observables, except for the nearest-neighbor
%density-density correlator which prefers a value of $U^{*}/t \sim 0$. 
%In the limit that the model is perfect, all estimates must 
%yield the same value; the differences reflect an inadequacy 
%of the on-site Hubbard model in describing all the data. 
%This is evidence for the need for long-range interactions.

%The discrepancy could be attributed to 
%(1) errors in experimental energies used in fitting 
%(2) use of a long-range model for the interactions 
%and (3) assumption that $t$ is not renormalized 
%on going from a long range to short range model. 
%Fig.~\ref{fig:models}(a) and (d) show the comparison 
%of the fitted energy, obtained by 
%taking the linear combination of the 
%{\it ab initio} VMC and DMC density matrices weighted by the optimized 
%parameters, and the corresponding measured energy. 
%A perfect agreement would correspond to all points falling exactly on the $y=x$ 
%line. By this measure, the Hubbard model is reasonable, but the presence 
%of outliers indicates the need for a more refined model.
%Thus, we explore the long-range model in the next section.
\begin{figure*}[htpb]
\centering
\subfigure[]{\includegraphics[width=0.48\linewidth]{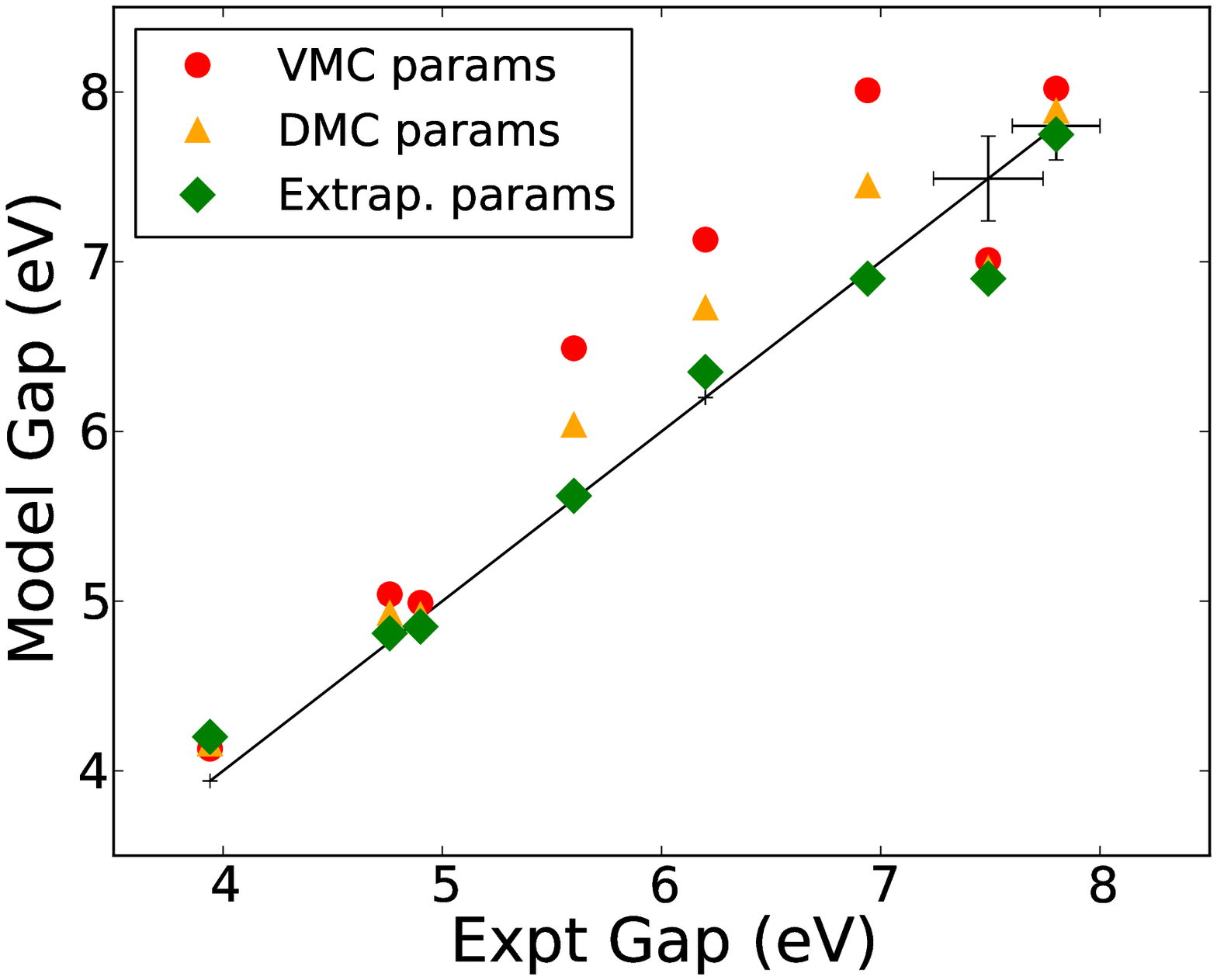}}
\subfigure[]{\includegraphics[width=0.48\linewidth]{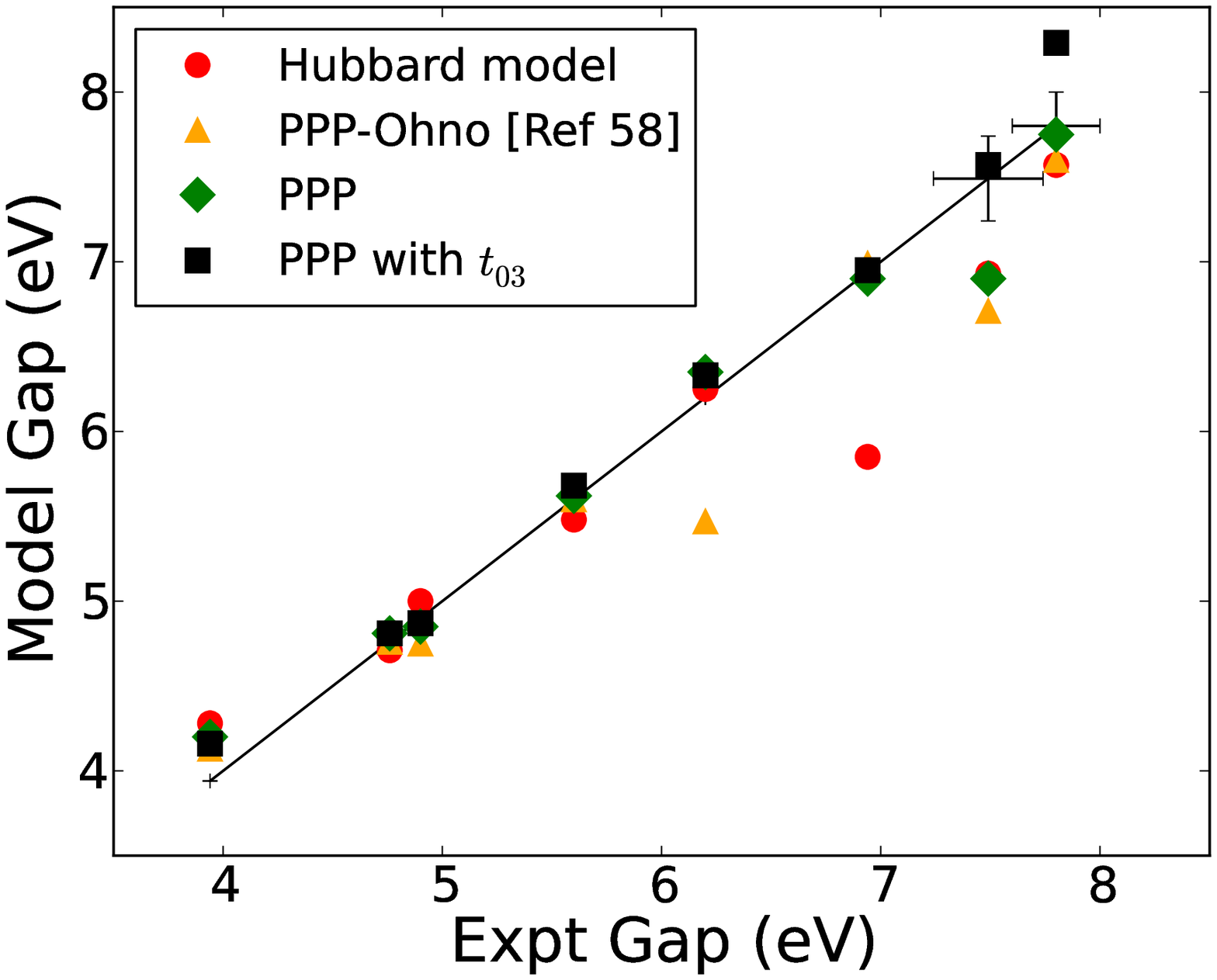}}
\caption{ \newt{Panel (a) shows the comparison of 
experimental energy gaps of the benzene molecule 
and energy gaps of the eigenstates of the 
extended Hubbard or Parisier-Pople-Parr (PPP) model 
on a six-site ring using optimized VMC, fixed-node DMC and extrapolated parameters, 
obtained from the N-AIDMD method. The extrapolated parameters 
remove, to a large extent, the bias from the other two calculations. 
All experimental values and associated error-bars 
are taken from Bursill et al. Ref.[\onlinecite{Bursill}], who 
used these values to fit to a PPP model with 
density-density interactions of the Ohno form in Eq.~\eqref{eq:Ohno}.
Panel (b) shows the comparison of experimental energy gaps of 
the benzene molecule and energy gaps of eigenstates for 
different model Hamiltonians. The on-site Hubbard model and the previous 
Ohno-parameterized Hamiltonian have at least one significant outlier, 
which is largely remedied by allowing all $V_{ij}$ and $U$ 
in the PPP model to be varied.  
}}
\label{fig:expt} 
\end{figure*}	

\begin{table*}[htpb]
\begin{center}
\begin{tabular}{|c|l|l|l|l|l|l|l|}
\hline
Parameter  &  PPP [Ref.~\onlinecite{Bursill}] & PPP-VMC & PPP-DMC & PPP-Extrap & $U^{*}$ [Ref.~\onlinecite{Schuler_graphene}] &  $U^{*}$-VMC & $U^{*}$-DMC \tabularnewline
\hline						   
\hline						   
 $t$        &   2.54    & $ 2.87(1)  $  &   $2.76(1) $  & 2.65(2)    &  2.54      &  $3.04(4)$  & $2.80(1)$ \tabularnewline
\hline				 
 $U$       &    10.06   & $ 11.95(4) $ &    $10.92(4)$  & 9.89(6)    &  3.04      &  $5.2(2)$   & $3.9(1)$ \tabularnewline
\hline				 
 $V_{01}$  &    7.18    & $ 7.47(5)  $ &    $7.13(3)$   & 6.78(8)    &   -         &    -               &        \tabularnewline
\hline				 
 $V_{02}$  &    5.11    & $ 5.40(3)  $ &    $5.41(2)$   & 5.40(4)    &   -         &    -               &        \tabularnewline
\hline				 
 $V_{03}$  &    4.57    &   4.57       &    4.57        & 4.57       &   -         &    -               &        \tabularnewline
\hline
\hline
 $U/t$     &    3.96    &   4.16(2)    &    3.96(2)     & 3.73(3)    &  1.20       &    1.7(1) & 1.39(5)   \tabularnewline
\hline
\hline
\end{tabular}
\caption{Model Hamiltonian parameters (in eV) from different downfolding methods, 
using data from states in the charge neutral sector of benzene. $V_{03}$ sets the 
constant shift or chemical potential for the 
interaction terms; its value has been set to match previous semi-empirical fits.}
\end{center}
\end{table*}
%In the first way, we consider a quantum state obtained from DMC
%with energy $\omega$ and ask what $U^{*}/t$ best describes 
%the double occupancy correlator at this spin state (remaining in 
%the same charge sector). In the second way, we assume that 
%the energy difference between this 
%state and the half filled ground state is described by a Hubbard model, whose 
%$U^{*}$ can be obtained completely from the knowledge of the density matrices

\subsection {Extended Hubbard model}
\label{sec:exthubb}
Having established the need for long-range interactions in benzene, 
we consider the extended-Hubbard 
or Parisier-Pople-Parr (PPP) model, 
\begin{equation}
H = -\sum_{ij} t_{ij} c_{i,\sigma}^{\dagger} c_{j,\sigma} + \text{h.c.} + U \sum_{i} n_{i,\uparrow} n_{i,\downarrow}+ \sum_{ij} V_{ij} n_i n_j
\label{eq:PPP}
\end{equation}
where $U$ is the on-site Hubbard interaction and $t_{ij}$ and $V_{ij}$ 
are inter-orbital hopping and density-density interactions. 
We compare our results with Bursill et al~\cite{Bursill}, who considered only a nearest neighbor hopping and 
took $V_{ij}$ to be of the Ohno form~\cite{Ohno}
\begin{equation}
	V_{ij}= \frac{U}{\sqrt{1 + (\alpha r_{ij})^{2}}}
\label{eq:Ohno}
\end{equation}
where $\alpha$ is a fit parameter and $r_{ij}$ is the spatial 
separation between nuclei. This parameterization 
has been widely used in the modelling of various organic polymers. 
Here do not make any assumptions about the form of the interactions 
and instead use the N-AIDMD method to determine these parameter values.

%More recently Schmalz~\cite{Schmalz} has discussed 
%the accuracy of the Mataga-Nishimoto (MN) form, 
%$V_{ij}= \frac{U}{1 + \alpha r_{ij}}$. 
%Both forms insure that in the large $r_{ij}$ limit $V_{ij} \rightarrow \frac{1} {r_{ij}}$. 
%The optimal parameters for these models are obtained by fitting 
%their spectra to available experimental energies.
%However, the results from both models differ - 
%the Ohno-form is reported to be $U=10.06 $ eV while 
%the MN form gives $U=8.32$ eV. 
We repeat analyses similar to those for the Hubbard model, in addition
to carefully looking at the variations in the 1- and 2-RDM matrix 
elements. This data has been discussed as part of Appendix B 
and has been shown in Fig.~\ref{fig:variations}. 
For a parameter to be reliably 
estimated there should be a large variation 
in the corresponding density matrix element 
for different wavefunctions in the low energy space. 
By this metric, we find that the next nearest neighbor hopping $t_{02}$ is irrelevant in the charge-neutral sector. 
We thus attempt to fit to a model only with the nearest neighbor 
$t \equiv t_{01}$ along with $U$, $V_{01}$ and $V_{02}$; 
$V_{03}$ is not needed as it simply sets the chemical potential. 

%Fig.~\ref{fig:models} shows energy fits to the on-site 
%and extended-Hubbard models both with VMC and DMC calculations. 
The inadequacies of the on-site Hubbard model, 
shown in Figs.~\ref{fig:models}(a) and (d),
are rectified by the extended one, 
shown in Figs.~\ref{fig:models}(b) and (e);
the maximum energy errors of about $2$ eV are reduced to $\approx 0.3$ eV. 
The root mean square errors are much smaller as well, 
reducing from $0.5$ eV to about $0.06$ eV. 
Adding the next-next nearest neighbor hopping $t_{03}$, 
shown in Figs.~\ref{fig:models}(c) and (f), 
only marginally improves the accuracy of the fit; $t_{03}$ 
is found to be only about a tenth of the value of $t_{01}$ 
suggesting that its effects can be largely accounted for by $t_{01}$.  
%We find that even in the $12$ eV window the maximum error 
%between the DMC and fitted energies 
%is now reduced to 0.15 eV shown in Figs~\ref{fig:expt}(b)

To assess the accuracy of these parameters, we 
compare the results of the model with experimentally 
available energies. First, as Fig.~\ref{fig:expt}(a) 
shows for the extended-Hubbard or PPP model, the extrapolated parameters 
give an improved agreement with experiment 
compared to the VMC or FN-DMC parameters. Most 
energy gaps of this model are in excellent 
agreement with available experimental energies, the errors are 
$0.2$ eV or less. The largest outlier at about $7.5$ eV 
is within $2\sigma$ of the experimental result. 

Next, we compare the experimental energies with energy gaps 
from various model Hamiltonians. While the 
Hubbard and the Ohno parameterizations reproduce most experimental 
gaps, especially at low energies, they have at least one significant outlier.
These outliers are correctly accounted for by the 
fitted PPP Hamiltonian, and improved upon by the introduction of $t_{03}$. 
Owing to the small value of $t_{03}$, more data in the N-AIDMD method 
may be needed to precisely estimate its value. The extreme sensitivity 
of the high energy eigenstates to $t_{03}$ may explain the deviation 
of the model gap from the experimental one at about $7.8$ eV.

We now discuss our parameter values and the errors associated with them; 
these have been summarized in Table I. 
While our PPP parameters are generally consistent with the Ohno form, 
there are some differences of the order of $0.3- 0.5$ eV, 
that improve the quality of the fitted energies. We emphasize that we 
have not provided any experimental inputs; rather we have used only the QMC data 
(energies and density matrices) from multiple states to obtain the 
Hamiltonian parameters. 

In order to check the robustness of the fit, 
we estimated errors in our parameters from a Jackknife analysis. In this scheme 
half the input states to the N-AIDMD method 
were randomly discarded and the fit performed with the retained half. 
Many such randomly generated ensembles of input data were taken and the resultant 
parameters were averaged over all of these. The difference of the parameters 
of this reduced data set and those obtained from the full data set provides 
an estimate of the systematic error, which we report in table I.
The other source of error is from statistical noise, 
which in the present case was found to be much smaller than 
the systematic error. 

\section{Application to a solid: Graphene}
As an application of the AIDMD methods to solid materials, we consider graphene, 
a 2D solid of carbon atoms arranged on a honeycomb lattice. 
Graphene has great potential technological applications, 
which has spurred much work devoted to understanding it thoroughly~\cite{graphene_review}.
In addition, there have been several proposals for engineering exotic phases 
in graphene~\cite{Herbut, Duric, Ghaemi}.
That said, it is only recently that systematic studies to estimate the role of 
electron-electron interactions~\cite{Wehling_graphene,Schuler_graphene,Abbamonte} 
have been carried out. While some of its long-distance properties appear to be adequately 
described by a tight-binding model, the short range features, 
crucial for phenomena such as magnetism, require more refined modeling.

\begin{figure}[htpb]
\centering
\includegraphics[width=1\linewidth]{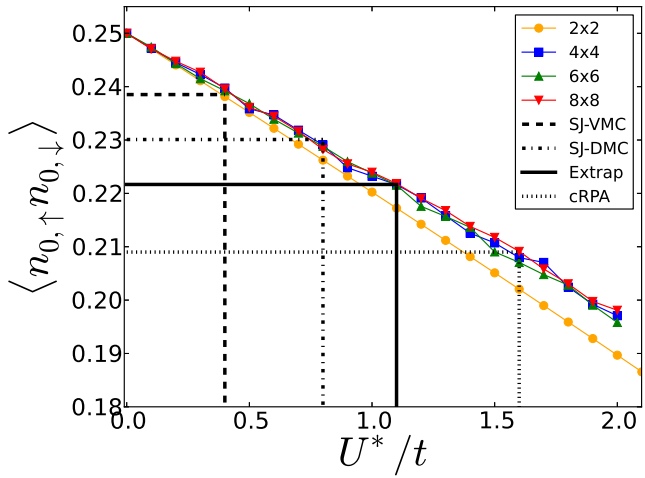}
\caption{Double occupancy correlator of a single $\pi$ orbital of the 
$L \times L$ ($L=2,4,6,8$) honeycomb lattice Hubbard model 
with periodic boundary conditions, as a function of $U^{*}/t$, 
computed in the half-filled singlet ground 
state. Comparisons are made with the values from variational (VMC), 
mixed (DMC) and extrapolated (Ext) estimators obtained from 
{\it ab initio} QMC calculations on $4 \times 4$ graphene 
using an optimized Slater-Jastrow (SJ) trial wavefunction. 
The cRPA estimate from Ref.[\onlinecite{Schuler_graphene}] is also shown.}
\label{fig:doubleocc} 
\end{figure}

\begin{figure*}[htpb]
\centering
\includegraphics[width=\linewidth]{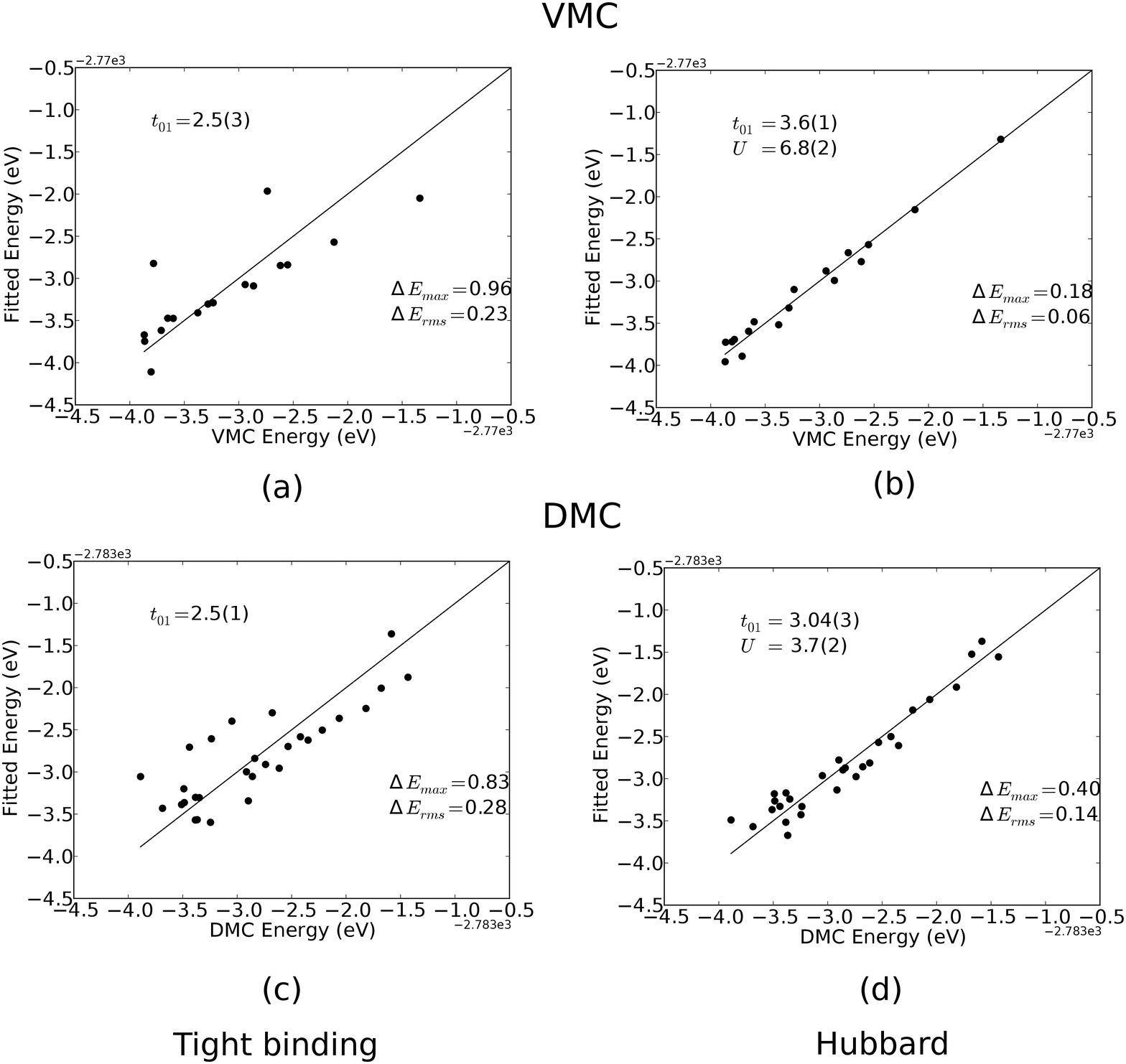}
\caption{\newt{Comparison of {\it ab initio} ($x$-axis) and fitted energies ($y$-axis) 
of the $3 \times 3 $ periodic unit cell of graphene, 
using the N-AIDMD procedure. The non-eigenstate data is generated by considering 
various spin excitations and nodal surfaces, all in the same charge sector 
(72 electrons). The {\it ab initio} energy is directly sampled using quantum Monte Carlo methods - 
for the variational Monte Carlo method (VMC) it corresponds to  
$\langle \psi_T | H | \psi_T \rangle / \langle \psi_T|\psi_T \rangle $ 
and for the fixed-node diffusion Monte Carlo 
(DMC) to $\langle \psi_T | H | \psi \rangle/ \langle \psi_T|\psi \rangle$ 
where $\psi_T$ and $\psi$ correspond to trial and projected wavefunctions 
respectively. 
The fitted energy is obtained from the optimized parameters by multiplying them 
by the corresponding {\it ab initio} density matrix elements. 
The top panels (a),(b) show the VMC results and the bottom ones 
(c),(d) show the fixed-node DMC ones. 
(a) and (c) correspond to the tight binding model 
on a honeycomb lattice, and (b) and (d) correspond to the on-site Hubbard model.
}}
\label{fig:models_graphene} 
\end{figure*}	

\begin{figure*}[htpb]
\centering
\subfigure[]{\includegraphics[width=0.48\linewidth]{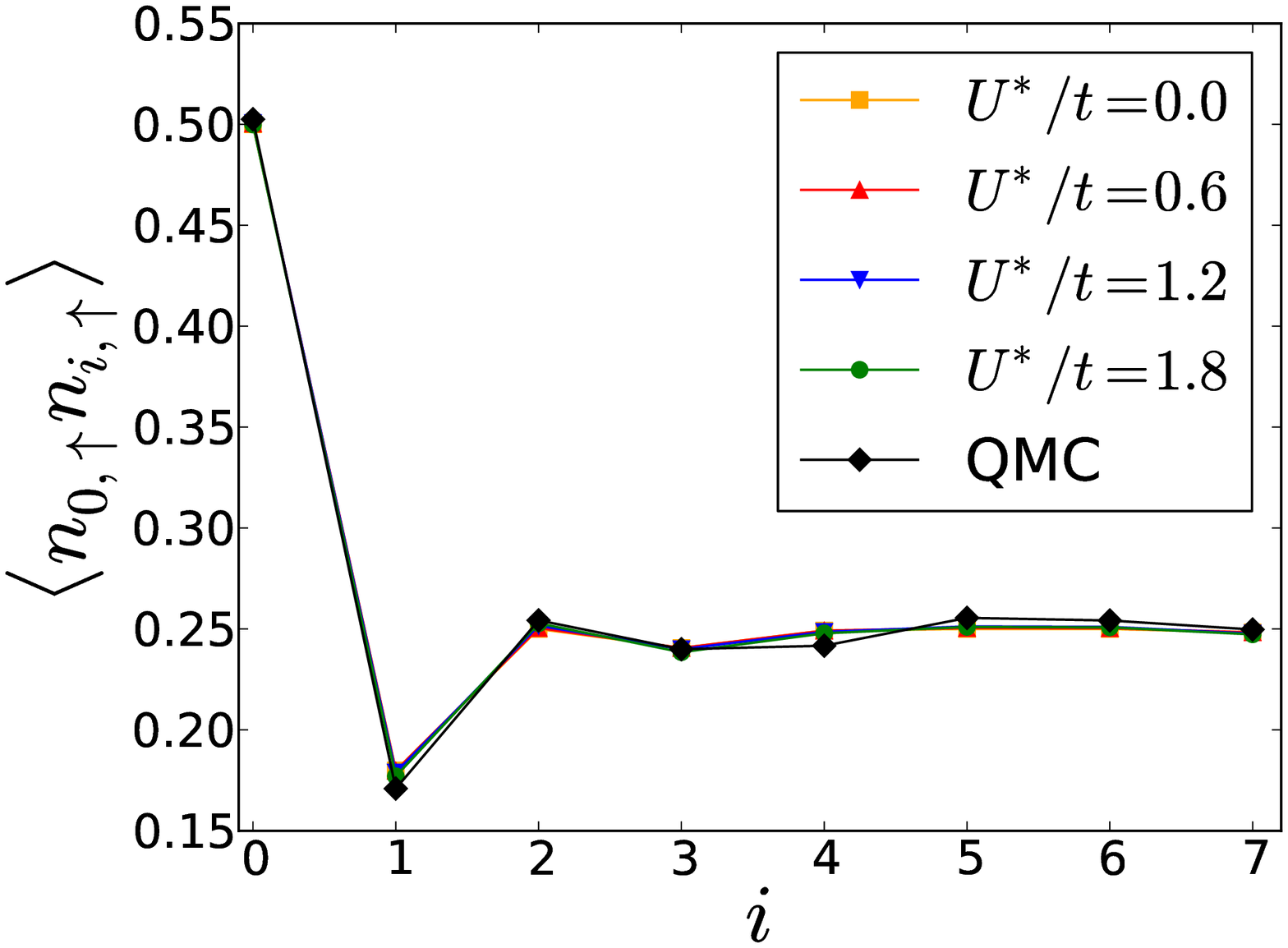}}
\subfigure[]{\includegraphics[width=0.48\linewidth]{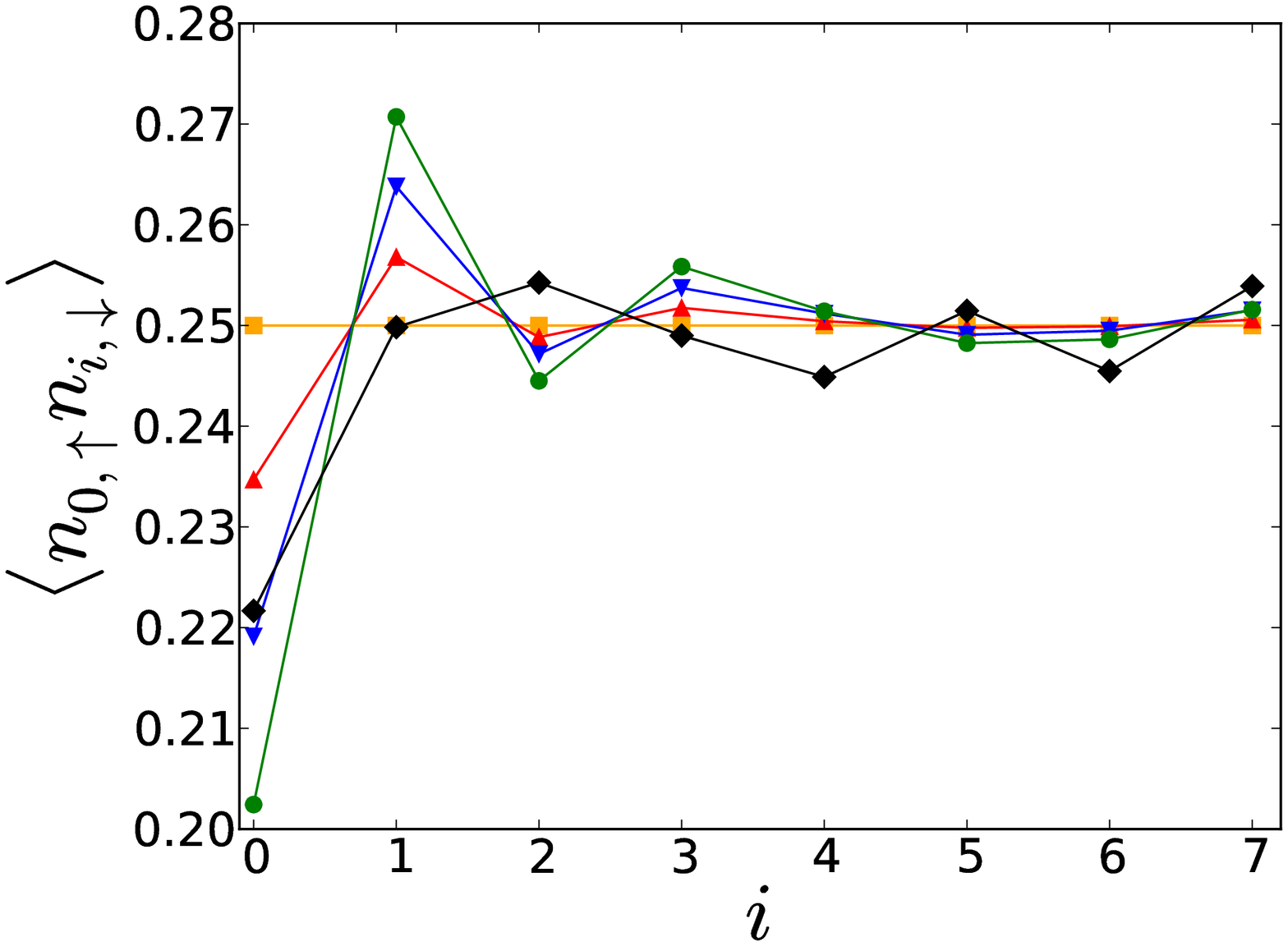}}
\caption{ \newt{Comparison of correlation functions of the 
half-filled, unpolarized ground state of $4 \times 4$ periodic cell of graphene 
from {\it ab initio} quantum Monte Carlo with those 
from determinantal quantum Monte Carlo calculations of the honeycomb 
lattice on-site Hubbard model for various values of $U^{*}/t$. 
Panel (a) shows the up density - up density correlator 
$\langle n_{0,\uparrow} n_{i,\uparrow} \rangle
\rangle$ and (b) shows the up density - down density correlator 
$\langle n_{0,\uparrow} n_{i,\downarrow} \rangle$ for the $i^{th}$ neighbor of 
a reference site (labelled $0$). The like-density correlators depend 
weakly on $U^{*}/t$ on the scale shown for all $i$. The correlator $\langle n_{0,\uparrow} n_{0,\downarrow} \rangle$ 
suggests $U^{*}/t \approx 1.2$, but the deviation for $\langle n_{0,\uparrow} n_{1,\downarrow} \rangle$ 
indicates that the Hubbard model overestimates the nearest-neighbor attraction between electrons of opposite spins.}}
\label{fig:corrs_graphene} 
\end{figure*}	

Early studies modelled graphene as a honeycomb lattice Hubbard model 
with a $U/t$ estimated to be $ \approx 3.8$. This would put graphene on the verge of a 
metal-insulator transition~\cite{Sorella_Tosatti, Sorella_Nature}. 
However, recent results by Wehling et al. realized the importance of 
long range interactions~\cite{Wehling_graphene},
which renormalize the on-site interaction to an effectively lower value. 
Schuler et al.~\cite{Schuler_graphene} report the effective $U^{*}/t$ 
to be $1.6 \pm 0.2$, which means graphene lies well in the 
semimetal phase of the honeycomb lattice Hubbard model. 

%The approach for downfolding from a lattice model with long range interactions (Hamiltonian $H$) 
%to that with an on-site $U^{*}$ (Hamiltonian $H^{*}$) 
%used the Peierls-Feynman-Bogoliubov variational principle, 
%which best matches the partition functions of the latter to the former. 
%This principle is closely related to matching the reduced density matrices of the ground state 
%in the zero temperature limit, albeit with subtle differences. 
%In particular, mapping one lattice model to another one is not what we refer to technically 
%refer as "downfolding" since there is no reduction of the number of electrons. 
%It must be emphasizes that our many body QMC calculations included all the $s$-like, 
%$\sigma$ and $\pi$ electrons of carbon. 
%The model Hamiltonian was downfolded onto the $\pi$-like orbital space.
Setting aside the question of determining all the long range interactions
in graphene, we ask what $U^{*}/t$ best describes our ground state QMC data. 
To do so, we first generated the $\pi$-like Wannier functions 
within QWalk~\cite{QWALK}, a representative of which has been shown in Fig.~\ref{fig:orbs}. 

Just as in the case of benzene, we used the fact 
that the effective strength of the Coulomb interaction 
$U^{*}/t$ is most sensitive to the 2-RDM element 
$\langle n_{i,\uparrow} n_{i,\downarrow}\rangle$. For the $4\times4$ 
unit cell with periodic boundary conditions, and using optimized Slater Jastrow wavefunctions, 
the extrapolated value is found to be $0.221(5)$ corresponding to a $U^{*}/t \approx 1.1(1)$.
This estimate of $U^{*}/t$ is obtained from comparisons to 
lattice determinantal QMC calculations, which were carried out for 
sizes ranging from $2 \times 2$ to $8 \times 8$ to check for finite-size effects. 
As Fig.~\ref{fig:doubleocc} shows, the $2 \times 2$ unit cell is distinctly different from 
the larger unit cells and the finite-size errors in the 
double occupation correlator are negligible beyond sizes $L\geq 4$. 
The finite size effects for other short range correlation functions (not shown) are also negligible beyond 
$L \geq 4$.

We also calculated many non-eigenstates for $3 \times 3$ graphene in an energy window $3$ eV 
above the ground state. Fig.~\ref{fig:models_graphene} shows the VMC and FN-DMC energy fits to 
tight-binding and Hubbard models. The hopping $t$ for the 
tight-binding model is found to be in the range of $2.2$ to $2.8$ eV, 
\newt{as is indicated in Figs.~\ref{fig:models_graphene}(a) and (c).
However,} a precise estimate of this parameter is not particularly meaningful because 
the model is inadequate at capturing many states, particularly spinful excitations, 
in this energy window. We note that a value of $t=2.80$ eV~\cite{Schuler_graphene} 
has been previously calculated with DFT methods. 

\newt{Figs.~\ref{fig:models_graphene}(b) and (d) show that} 
the Hubbard model reduces the errors of the tight-binding model, 
and the value of $U^{*}/t$ is found to be in the range of $1.9$ (VMC) to $1.3$ (FN-DMC). 
The latter estimate is expected to be more accurate and hence closer to the 
true value of $U^{*}/t$ in a small energy window 
associated with the ground state. We also note that this value 
is within $2\sigma$ of the value 
derived from the constrained RPA parameters~\cite{Schuler_graphene}. 
However the Hubbard model too has outliers of about $0.4$ eV, 
which are large for an accurate model of a solid material. 

This inadequacy is confirmed by assessing various correlators in the half filled ground state. 
Fig.~\ref{fig:corrs_graphene} shows the up density-up density and 
up density- down density correlators as a function of distance 
between carbon atoms. On the scale of Fig.~\ref{fig:corrs_graphene}(a) 
(and well within the accuracy of our calculations) the like spin correlations were 
captured well for all values of $U^{*}/t$ in the range from $0$ to $2$. However, as 
Fig.~\ref{fig:corrs_graphene}(b) shows, the Hubbard model for large $U^{*}/t$ 
tends to exaggerate the the effective interaction between the two 
electron spin flavors at small separations. In particular, the 
nearest neighbor unlike spin density-density correlator is found 
to be in better agreement with $U^{*}/t \sim 0$ than any finite value. 
This, just like the case of benzene, suggests the need 
for longer range interactions in the model. There are also small 
deviations between the {\it ab initio} QMC and the Hubbard model results 
at longer distances. These correlations do 
not depend significantly on $U^{*}/t$ and these data in isolation do not rule out any model.
 
\section{Conclusion} 
We have demonstrated {\it ab initio} density matrix downfolding (AIDMD) methods 
where {\it ab initio} quantum Monte Carlo (QMC) 
data is used to fit simple effective Hamiltonians. We have elaborated 
on the fitting procedures and the intricacies of the QMC method 
needed to perform calculations. The limitations of the model 
were judged by assessing the quality of the fitted energies and 
2-body density matrices. This feature is useful for constructing 
refined models needed for the accurate simulation of real materials. 

For the benzene molecule, while the on-site 
Hubbard model with $U^{*}/t \approx 1.2 \pm 0.2$ 
was able to capture most features of the QMC ground state data, 
the deviations of the density matrices revealed the need for 
longer range interactions. \newt{Including these interactions improved the agreement 
of the model with both the QMC results and the experimental data.} 
This effective Hamiltonian parameterization could be used to calculate 
low-frequency response functions and to check semi-empirical methods.

Since QMC calculations use size-consistent wavefunctions for extended systems 
and scale favorably, we believe the type of calculations presented 
here will be a promising alternative to DFT-based downfolding approaches 
for solid materials. Our demonstration for the single band model of graphene 
yielded an effective $U^{*}/t=1.3\pm0.2$, in the same 
range as a recently reported estimate based on the constrained-RPA 
method~\cite{Schuler_graphene}. \newt{We leave a more 
detailed characterization of interactions in graphene to future work.}

In more complicated materials, 
where the form of the Hamiltonian is unclear, we suggest that 
the dominant terms can be obtained from canonical transformation 
theory followed with an accurate fit to the QMC data. 
\newt{In this spirit}, it will also be useful to compare the predictions of the proposed 
AIDMD schemes with other complementary proposals for downfolding~\cite{Zhang, Zgid, Watson_Chan}. 

Finally, we remark that previously unsolved model Hamiltonians 
are now being accurately treated with tensor 
network methods~\cite{Stoudenmire_White_2D_DMRG, Corboz_Rice_Troyer}. 
With parallel advances in the {\it ab initio} QMC simulation of high-temperature 
superconductors~\cite{Wagner_cuprates, Krogel_cuprates}, a clear future direction 
is to deduce more refined models for these compounds, using the ideas discussed in the paper. 

\section{Acknowledgement} 
We thank David Ceperley, Cyrus Umrigar, Garnet Chan, Shiwei Zhang, 
Steven White, Gabriel Kotliar, Bryan Clark, Norman Tubman, Victor Chua 
and Miles Stoudenmire for discussions. 
We would also like to thank Cyrus Umrigar for running very useful checks against 
his quantum Monte Carlo code (CHAMP) in the early stages of 
this work. We acknowledge support from SciDAC grant DE-FG02-12ER46875. 
\newt{Computational  resources were provided by the DOE INCITE SuperMatSim program and 
the Taub campus cluster at the University of Illinois at Urbana-Champaign.} 

\section{Appendix A: Estimation of true effective Hamiltonian 
parameters from VMC and FN-DMC calculations} 
\label{sec:app}

In section~\ref{sec:qmc_param}, we discussed 
the discrepancy between parameters obtained from VMC and FN-DMC. 
This is attributed to the inability of the trial wavefunctions 
used in VMC to provide an accurate description of the core and virtual spaces.  

As mentioned in the text, ideally, only the FN-DMC calculations 
should be used to estimate the parameters. However, 
the mixed estimator in FN-DMC is biased because 
of the inaccuracy of the trial wavefunction. 
To minimize this bias, we propose estimators 
which combine FN-DMC and VMC parameters.

Consider a trial VMC wavefunction $\psi_T$ 
which deviates from the DMC wavefunction $\psi$ by a small 
amount $\delta \psi_h$ orthogonal to $\psi_T$ i.e., 
\begin{equation}
	| \psi \rangle = |\psi_T \rangle + \delta |\psi_h \rangle
\end{equation}

To obtain the parameter $p$ coupled to an operator $\rho_p$ in the effective Hamiltonian i.e. 
\begin{equation}
	H = \sum_{p} p \;\; \rho_p
\end{equation}
we take partial derivatives with respect to the change of a 
density matrix element i.e.
\begin{equation}
	p = \frac{\partial \langle \psi | H | \psi \rangle} {\partial \langle \psi | \rho_p | \psi \rangle}
\end{equation}
Since the pure estimators for projected (DMC) wavefunctions 
are not easily evaluated in QMC, we use other estimators using which we indirectly 
obtain $p$. 

The parameter obtained from the mixed estimators within FN-DMC, is formally defined as,
\begin{equation}
	p_{D} = \frac{\partial \langle \psi | H | \psi_T \rangle} {\partial \langle \psi | \rho_p | \psi_T \rangle}
\end{equation}
%This FN-DMC estimator for the parameters is biased, as it involves the VMC 
%wavefunctions. As discussed previously, VMC does not adequately account for 
%the ground state configuration of the core and virtual electrons. Thus, we propose a 
%a scheme that minimizes this bias. 
On substituting the relation between $\psi$ and $\psi_T$ wavefunctions, 
we get,
\begin{align}
	p_{D} = \left( \frac{\partial \langle \psi | H | \psi \rangle} {\partial \langle \psi | \rho_p | \psi \rangle} \right) \left(1-\delta \frac{\partial \langle \psi| H | \psi_h \rangle}{\partial \langle \psi | H | \psi \rangle } \right) 
\left( 1 - \delta \frac{\partial \langle \psi| \rho_p | \psi_h \rangle}{\partial \langle \psi| \rho_p | \psi \rangle} \right)^{-1}
\end{align}
which to linear order in $\delta$ is,
\begin{align}
	p_{D} \approx p \left(1 + \delta \left(  \frac{\partial \langle \psi| \rho_p | \psi_h \rangle}{\partial \langle \psi | \rho_p | \psi \rangle } -  \frac{\partial \langle \psi| H | \psi_h \rangle}{\partial \langle \psi | H | \psi \rangle } \right) \right)
\label{eq:pdmc_approx}
\end{align}

A similar expression for parameters obtained from VMC calculations, 
\begin{equation}
	p_{V} = \frac{\partial \langle \psi_T | H | \psi_T \rangle} {\partial \langle \psi_T | \rho_p | \psi_T \rangle}
\end{equation}
is derived and leads to the result, 
\begin{equation}
	p_{V} \approx p \left(1 + 2 \delta \left(  \frac{\partial \langle \psi| \rho_p | \psi_h \rangle}{\partial \langle \psi | \rho_p | \psi \rangle } -  \frac{\partial \langle \psi| H | \psi_h \rangle}{\partial \langle \psi | H | \psi \rangle } \right) \right)
\label{eq:pvmc_approx}
\end{equation}
where we have used the hermiticity of the Hamiltonian $H$ and the operator $\rho_p$. 

On combining equations~\eqref{eq:pdmc_approx} and \eqref{eq:pvmc_approx}, to leading order in $\delta$ we get, 
\begin{equation}
	p = 2 p_{D} - p_{V} + O ({\delta}^2)
\end{equation}
%An alternate way of extrapolating the true parameters is,
%\begin{equation}
%	p = \frac{p_{D}^{2}} {p_{V}}
%\end{equation}
As expected, all estimates are consistent with each other 
in the limit of exact wavefunctions. 
%in the limit that the deviations (order $\delta$) 
%between the VMC and DMC parameters is small. 

\section{Appendix B: QMC data for the benzene molecule} 
\label{sec:app}

Table II shows the QMC data for several eigenstates of benzene. 
Most of these states along with many other non-eigenstates constitute our data set for fitting a model. 

\begin{table*}[htpb]
\begin{center}
\begin{tabular}{|c|c|c|c|c|c|c|c|}
\hline
Spin &   DFT      &   SJ-VMC      &  SJ-DMC	 &  CISDJ-VMC    &  CISDJ-DMC  & $N_\uparrow,N_\downarrow$ & Used in Fit?\tabularnewline 
\hline     					      
\hline     					      
 0   &   -37.6303   & -37.6229(6)   & -37.7213(9)	 & -37.6352(6)   & -37.7259(9) & 2.96,2.96  & Yes \tabularnewline
\hline     					      
 1   &   -37.4634   & -37.4546(6)   & -37.5555(7)	 & -37.4814(6)   & -37.5707(7) & 3.94,1.98  & Yes \tabularnewline
     &              &               &          	         & -37.4561(6)   & -37.5479(6) & 3.94,1.98  & Yes \tabularnewline
     &              &               &       	         & -37.4531(6)   & -37.5470(6) & 3.94,1.98  & Yes \tabularnewline
\hline     					      
 2   &   -37.3203   & -37.2987(6)   & -37.3974(7)	 & -37.3141(6)   & -37.4020(7) & 4.92,1.00  & Yes \tabularnewline
\hline     					      
 3   &   -37.0378   & -37.0116(4)   & -37.1074(7)	 & -37.0118(4)   & -37.1083(7) & \color{RED} {4.88,0.02}  & No \tabularnewline
\hline     					      
\hline 
\end{tabular}
\caption{Energy of different spin eigenstates in the charge-neutral sector of benzene 
from DFT and QMC methods. SJ refers to the Slater Jastrow wavefunctions. CISDJ 
refers to multi-determinantal Jastrow wavefunctions where the determinants are obtained from a CI-singles doubles 
calculations and their weights optimized within VMC. All DMC calculations used a time step of 0.01 Ha$^{-1}$. 
The states that had significantly different occupations from 
the expected values were not used in the fitting as they can not be 
described by an effective Hamiltonian involving only six $\pi$ orbitals.}
\end{center}
\end{table*}

The calculations confirm the general expectation that significant energy gains are obtained by 
improving wavefunctions going from the single-Slater-Jastrow form to the 
multi-determinantal-Jastrow form (the determinants being selected 
from a CISD calculation). Moreover, the DMC calculations 
improve total energies significantly; typically, 
the DMC values are $2-3$ eV lower than the corresponding VMC value.

The total electron count from the one-body density matrix is 
assessed to verify the validity of fitting to a six-$\pi$ orbital Hamiltonian. 
Table II shows that for charge-neutral benzene, 
the singlet state ($S=0$) has up and down electron 
counts of 2.96 each, which are close to the expected values of 
3,3. For the $S=2$ state, roughly $9$ eV above the ground state, 
the deviations were slightly larger; the summed 
occupation numbers were 4.92 and 1.0 in comparison to the 
expected values of 5 and 1. 
Since there is a slight deviation of these numbers from 
integers, we rescale the one and two body density matrices 
by factors (all slightly greater than 1) that correctly 
accounts for sum rules for each individual state used in the AIDMD 
methods. 

However, for the $S=3$ state, the electron 
occupation numbers deviate significantly 
from the corresponding value in the model; 
almost by one integer. This indicates that the $S=3$ state is 
inadequately described by the proposed Hamiltonian and hence 
can not be used in the fitting procedure. 
This deviation is not completely unexpected since this state is 
$\sim 17$ eV above the ground state and a potential high-energy state. 
Said differently, the active space at this energy scale is considerably 
different from that assumed for the ground state and its low-energy excitations.
Thus, this QMC data suggests that it is only reasonable for the effective 
Hamiltonian concept to hold only in an energy window of the order of $10 $ eV above 
the ground state.

We now assess some aspects of our non-eigenstate data. 
\newt{As mentioned in the main text, heuristics were used to 
construct these states. For example, for every near-eigenstate 
in a symmetry sector that was represented by multi-determinantal Jastrow form, 
we changed the determinantal coefficients to generate new wavefunctions. 
We checked the 1-RDM to make sure that it had the correct 
total electron number (6 electrons) on the localized orbitals 
that constitute our active space. Moreover, if the change in determinantal coefficients 
led to an energy-average outside our pre-decided energy window, the new state 
was discarded from the N-AIDMD procedure.}

\begin{figure*}[htpb]
\centering
\subfigure[]{\includegraphics[width=0.24\linewidth]{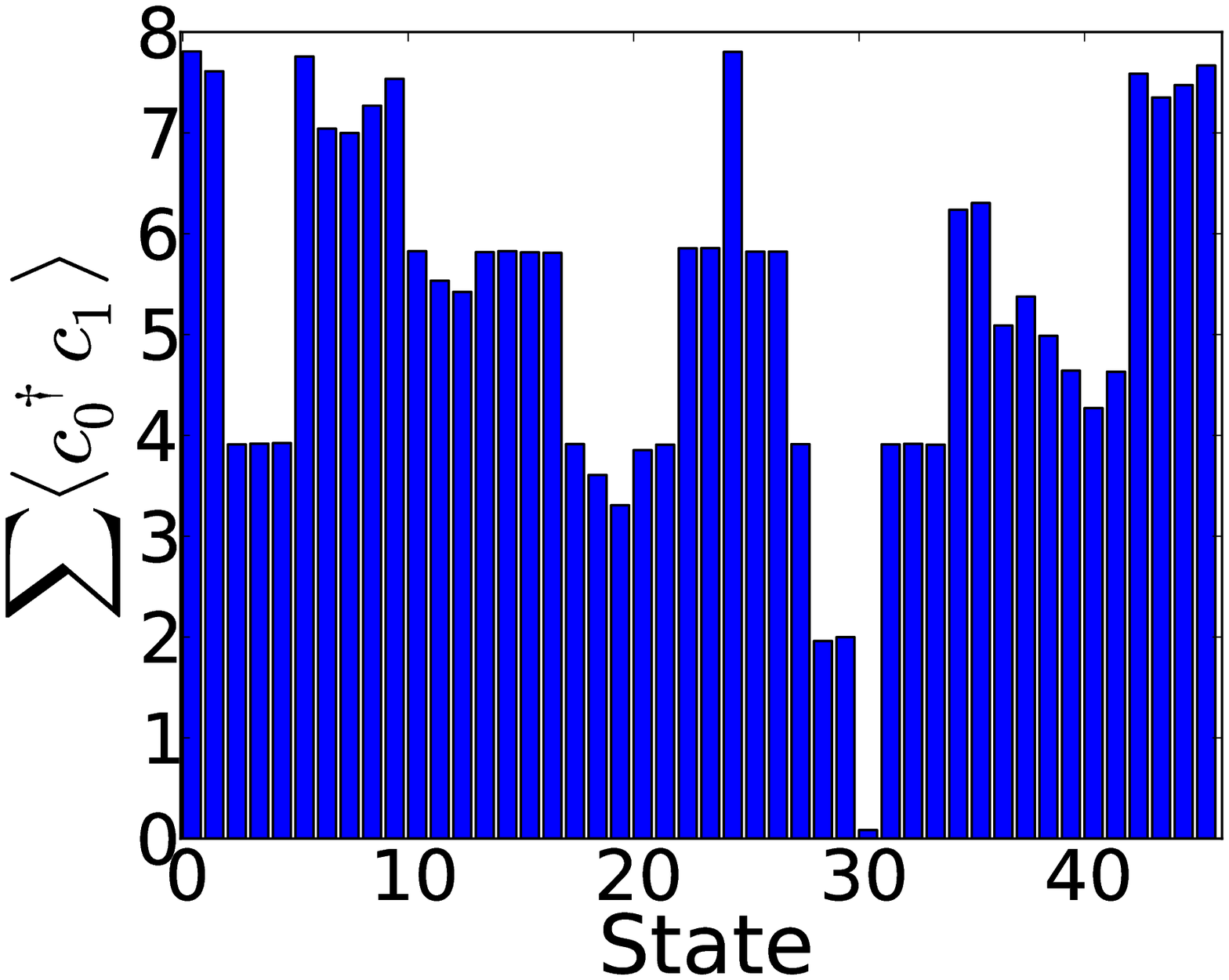}}
\subfigure[]{\includegraphics[width=0.24\linewidth]{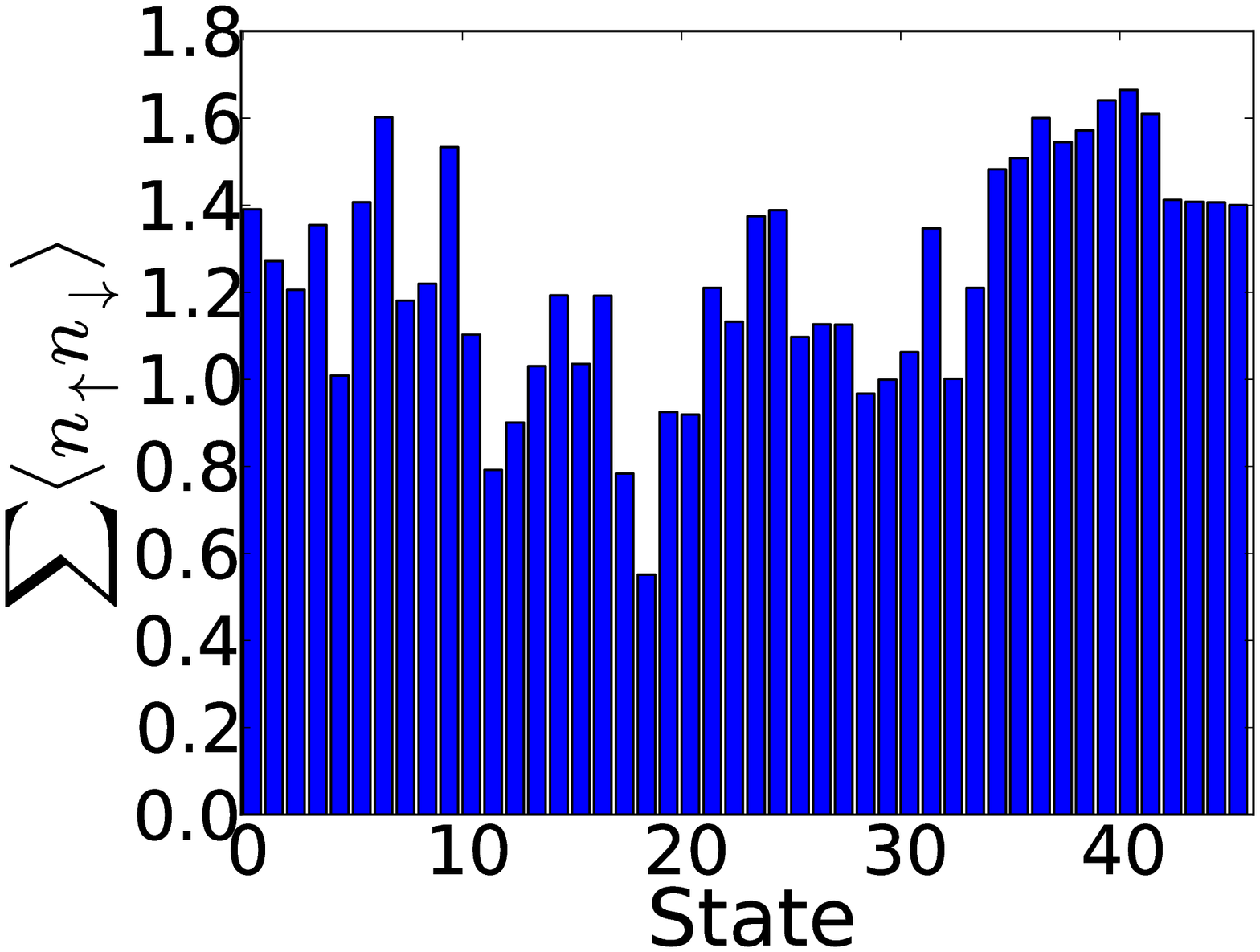}}
\subfigure[]{\includegraphics[width=0.24\linewidth]{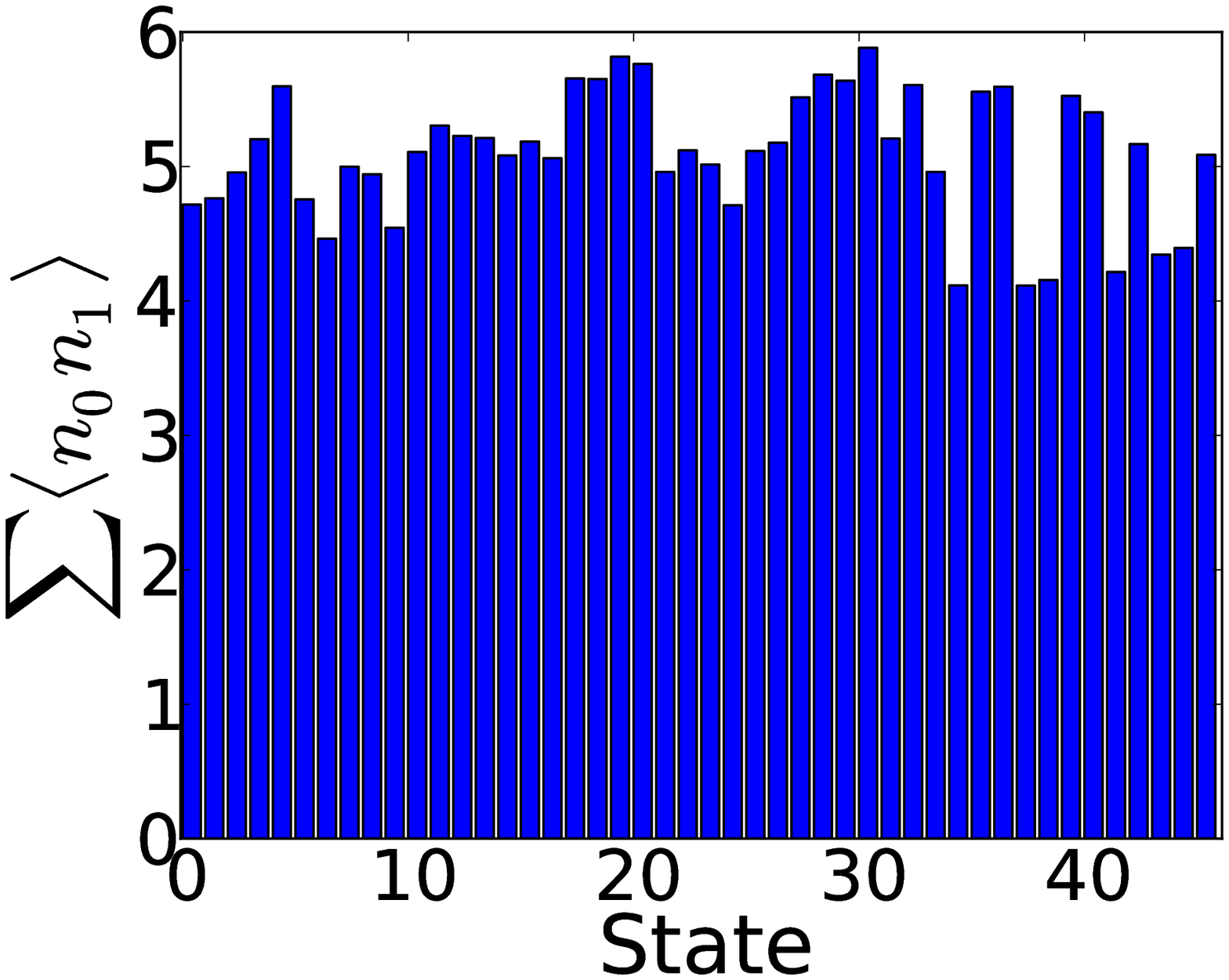}}
\subfigure[]{\includegraphics[width=0.24\linewidth]{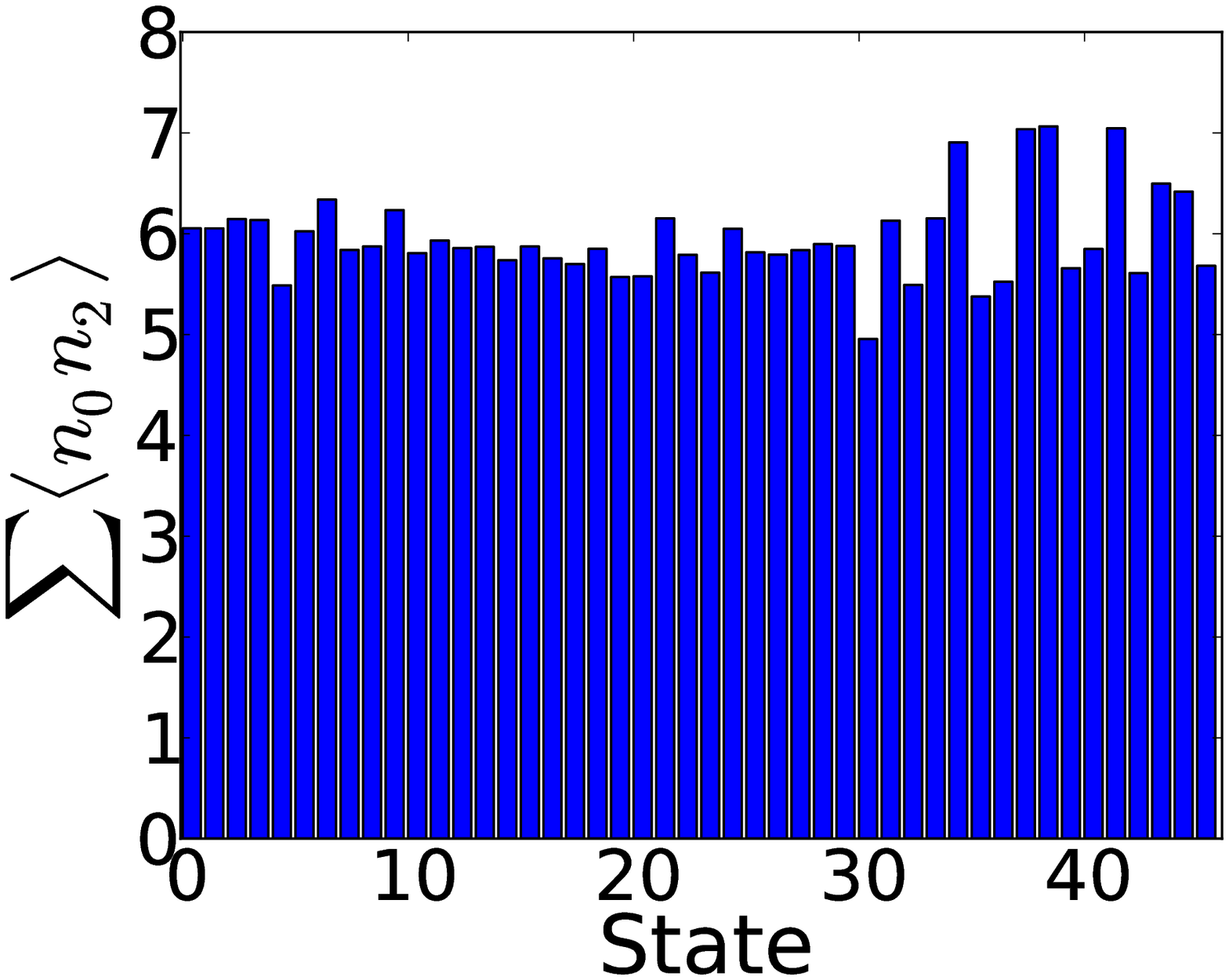}}
\caption{ \newt{Variation of density matrix elements for various states of the benzene molecule 
used in the N-AIDMD method.} The summation sign $\sum$ indicates a sum over all RDM elements that 
couple to the same parameter in the lattice model. The panels show (a) the sum over all the 
nearest neighbor one body density matrix elements summed over both spin types and (b) 
the sum over on-site double occupancies. (c) and (d) show sums over nearest and next-nearest 
neighbor density-density correlators respectively. As is discussed in the text, 
variations in the density matrix elements are needed for the corresponding 
parameters to be estimated, else they can be taken to be effectively zero.}
\label{fig:variations}
\end{figure*}

For the N-AIDMD method, we desire large variations in the 
density matrix elements for different 
wavefunctions in the low-energy space, in order to accurately 
estimate the Hamiltonian parameters. Fig.~\ref{fig:variations} 
shows these variations for all relevant density-matrix elements (within FN-DMC)
that were needed for estimating the parameters of the 
extended Hubbard model. $t_{02}$ was found to be irrelevant, 
as the summed density matrix elements coupling to it were found to not vary 
significantly (not shown in the plot).

\bibliographystyle{prsty}
%\bibliography{refs}

\end{document}